\documentclass[a4paper,usenatbib]{mnras}
\usepackage[T1]{fontenc}
\usepackage{ae,aecompl}

\usepackage{graphicx}
\usepackage{hyperref}
\usepackage{amsmath}
\usepackage{amssymb}
\usepackage{mathtools}
\usepackage{bm}
\usepackage{cleveref}

\title[Effects of general relativity on glitch amplitudes]
{Effects of general relativity on glitch amplitudes and pulsar mass upper bounds}

\author[ M.Antonelli, A. Montoli, \& P.Pizzochero]{
     M.~Antonelli$^{1,2}$ \thanks{E-mail: marco.antonelli@unimi.it}, A.~Montoli$^{1,2}$,
    P.~M.~Pizzochero$^{1,2}$\\ \\
    $^{1}$Dipartimento di Fisica, Universit\`a degli Studi di Milano, Via Celoria 16, 20133 Milano, Italy\\ 
    $^{2}$Istituto Nazionale di Fisica Nucleare, sezione di Milano, Via Celoria 16, 20133 Milano, Italy}
\begin{document}
    %
    %
    \pagerange{\pageref{firstpage}--\pageref{lastpage}} \pubyear{2016}
    \maketitle
    \label{firstpage}
    \begin{abstract}
Pinning of vortex lines in the inner crust of a spinning neutron star may be the mechanism that enhances the differential rotation of the internal neutron superfluid, making it possible to freeze some amount of angular momentum which eventually can be released, thus causing a pulsar glitch. We investigate the general relativistic corrections to pulsar glitch amplitudes in the slow-rotation approximation, consistently with the stratified structure of the star. We thus provide a relativistic generalization of a previous Newtonian model that was recently used to estimate upper bounds on the masses of glitching pulsars. We find that the effect of general relativity on the glitch amplitudes obtained by emptying the whole angular momentum reservoir is less than 30\%. Moreover we show that the Newtonian upper bounds on the masses of large glitchers obtained from observations of their largest recorded event differ by less than a few percent from those calculated within the relativistic framework. This work can also serve as a basis to construct more sophisticated models of angular momentum reservoir in a relativistic context. In particular, we present two alternative scenarios for  ``rigid'' and ``slack'' vortex lines, and we generalize the Feynman-Onsager relation to the case when both entrainment coupling between the fluids and a strong gravitational field are present.
\end{abstract}
\section{Introduction}
\label{sec:into}
Pulsars are extremely stable astronomical clocks. However, their slow and predictable 
secular spin-down, due to conversion of rotational energy into electromagnetic radiation and particle wind, can be  interrupted occasionally by impulsive spin-up events known as glitches [see e.g. \citet{espinoza2011} for a comprehensive observational study and \citet{haskell_review} for an up-to-date description of pulsar glitch models]. 

Large pulsar glitches, like the ones exhibited by the Vela, are usually interpreted as manifestations 
of the presence of quantized vortices which thread the internal superfluid bulk made of Cooper-paired neutrons \citep{ANDERSON1975}.
A central role in this scenario is played by the effectiveness of the pinning mechanism \citep{alpar77, Epstein1988}: superfluid vortices immersed in the solid nuclear lattice of the inner crust can find energetically convenient some particular vortex-nucleus
geometrical configurations \citep{Donati2004a,Donati2006}, which strongly hinder their motion. 
Also pinning to the magnetic flux tubes in the core is sometimes invoked, as this possibility opens interesting scenarios for both pulsar glitches and the study of the internal magnetic field evolution \citep{Ruderman1998}. The more the vortices are pinned, the more the superfluid is decoupled from the normal component and can flow independently from the spinning-down normal component. 

In the extreme case of perfect pinning, the only interaction between the two components is non-dissipative and consists of the so-called entrainment effect, early introduced to describe the dynamics of superfluid mixtures \citep{AB75}.
As a consequence of the density-dependence of both pinning and entrainment, the neutron superfluid develops a differential rotation with respect to the spinning-down normal component.
The velocity lag built up between the two components can also depend on how much vortex lines are able to ``creep-out'' the bulk superfluid; one of earliest models of this type is that of \citet{Alpar1984a}. The recoupling of the superfluid component following a large-scale vortex unpinning (triggered by a still unknown mechanism) leads to a small and fast 
spin-up of the normal component, namely a glitch \citep{epstein_baym92}. 

A consistent description of pulsar rotational dynamics was given by \citet{antonelli17}, based on the assumption of paraxial (rigid) vortices in a Newtonian framework; general relativity (GR) was only used to determine the stellar hydrostatic structure. This is not, however, an accurate description for neutron stars. For example, the most obvious correction that should be considered in GR is the use of relativistic moments of inertia (e.g. \citet{hartle67} and \citet{ravenhall_pethick94}). In the same work, the authors proposed a method to estimate the maximum glitch amplitude of a neutron star of given mass: thus, the largest glitch recorded in a given pulsar can  in principle put constraints on its (observationally unknown) mass, as recently proposed in \citet{pizzochero17}. Alternative methods for constraining the structure of neutron stars using glitch observations have already been carried out in previous works, among which \citet{datta93} and \citet{alpar+93} for the Vela pulsar as well as the more recent \citet{link99}, \citet{Andersson2012}, \citet{chamel13} and \citet{Ho2015}.

The aim of this work is  to embed in a general relativistic framework the method used to set an upper limit to the glitch amplitudes,
in order to estimate the corrections  due to the presence of a strong gravitational field  to the mass upper bounds presented in \citet{pizzochero17}.  Our study is developed in the slow-rotation approximation, appropriate for constructing models of non-millisecond pulsars in GR \citep{andersson_comer01}. 
The methodology is based on the two-fluid representation of superfluid neutron stars, wherein all the charged components 
(protons, electrons and lattice nuclei) are considered as a single fluid coexisting with the superfluid neutrons.
Following standard terminology \citep{AnderssonLivRev}, these two fluids are usually referred to as ``neutrons'' and ``protons''
and are coupled via the so called entrainment effect [see  \citet{chamelReview} for an introduction to entrainment in neutron stars].

The paper is organized as follows: in Section \ref{sec:newton} we recall some properties of the Newtonian model introduced in \citet{antonelli17}, adding a new local unpinning prescription for
vortex lines with negligible tension at the hydrodynamical scale, henceforth dubbed ``slack'' vortices. 
In Section \ref{sec:GR}, we generalize the previous results in the framework of a slowly rotating star in general relativity. Finally, in Section \ref{sec:numerical} we study the variation of the maximum glitch amplitudes under the different prescriptions, and we evaluate the difference in the inferred mass upper limit. The basic formalism needed here is briefly reviewed in the Appendix A, while in Appendix B we show a generalization of the Feynman-Onsager relation when a strong gravitational field is present. 
\section{Maximum glitch amplitudes: Newtonian framework}
\label{sec:newton}
An axisymmetric Newtonian model for pulsar glitches that is consistent with the stratified structure of a neutron star, the differential rotation of the superfluid 
and the presence of non-uniform entrainment and pinning is described in \citet{antonelli17}, hereafter Paper-I.
This model provides a generalization of the early two-rigid-components model of \cite{BAYM1969} and is rigorously built starting from some working hypotheses: 
the motion of matter is purely circular (null macroscopic meridional circulation) and the background stellar structure is that of non-rotating hydrostatic equilibrium in GR.
Physical motivation behind this choice is that one does not consider precession, nor rotational frequencies comparable to the mass-shredding limit. 

The normal component (labeled by $p$) rotates rigidly with a slowly decreasing angular velocity $\Omega_p$, 
while the superfluid component (labeled by $n$) can rotate non-uniformly with  angular velocity $\Omega_n=\Omega_{p}+\Omega_{np}$, but still around 
a common and fixed rotational axis.
Using standard cylindrical coordinates
    \footnote{
        In this work we use  standard spherical coordinates $(r,\theta,\varphi)$, where $\theta=0$ denotes the positive direction of the rotational axis and $\theta=\pi/2$ is the equatorial plane. The coordinate $\varphi$ is the azimuthal angle. 
        To follow the notation of Paper-I, we also use  cylindrical coordinates $(x,\varphi,z)$, defined as $x=r \sin\theta$ and $z=r \cos\theta$.
        The z-axis coincides with the rotational axis, whereas the inner-outer crust interface is the sphere of radius $R_d$, 
        namely $(x,\varphi,\pm z(x))$ with $z(x)=(R_d^2-x^2)^{1/2}$. The volume element in  flat space is $d^3\!x =dx d\varphi dz \, x$
        or $d^3\!x =drd\theta d\varphi \sin\theta r^2 $. 
        Within the relativistic slow-rotation formalism, the spherical coordinates are Schwarzschild-like coordinates: 
        in particular the coordinate $r=(x^2+z^2)^{1/2}$ represents the circumferential radius.
    },
the differential angular velocity lag between the components is a function $\Omega_{np}(x,z)$, which can vary in time but does not depend 
upon the coordinate $\varphi$, as can be shown by considering the continuity equation for the $n$-component.

Entrainment  provides a non-dissipative interaction between the two components; it is introduced in the model by following 
the Newtonian formalism presented in \citet{Prix2004}, where the momenta per baryon of each fluid $\bm p_n$ and $\bm p_p$  are linear combinations of the velocities of both fluids.
In particular, we are interested in the azimuthal component of the momenta
\begin{align}
& p_{ n \varphi } = m_n \, x \,(\Omega_p+(1-\epsilon_n)\Omega_{np}) 
\label{pnphi-newton}
\\ 
& p_{ p \varphi } = m_p \, x \,(\Omega_p+\epsilon_p\Omega_{np}) \, ,
\label{ppphi-newton}
\end{align}
the other components being zero. Here $m_n$ and $m_p$ represent the mass per baryon of the two fluids and the dimensionless entrainment parameters $\epsilon_n$
and $\epsilon_p$ obey the constrain $ m_n n_n \epsilon_n = m_p n_p \epsilon_p $, where $n_n$ and $n_p$
are the baryon number densities of each component. The total baryon density is $n_B=n_n+n_p$.
Since the superfluid is composed entirely of neutrons, we can consider $m_n$ to be the neutron bare mass.
The quantum of circulation around a single vortex line is thus $\kappa = h/(2 m_n)$, where $2m_n$ is the mass of a Cooper pair of neutrons.
Things are more subtle for the normal component, consisting of a neutral mixture of protons, leptons (in particular electrons), thermal excitations 
of the neutron superfluid and crustal neutrons that are not in the conduction band \citep{Carter2006}.
As a first approximation, $\beta$-equilibrium for a three-component star (electrons, protons and neutrons) tells us 
that $m_p$ is at least the sum of the proton  and  electron bare masses \citep{sourie_etal16}. 
Further contributions from the non-superfluid neutrons in the nuclear clusters can only rise the value of $m_p$ even closer to $m_n$.

In general, by using the constrain on the entrainment parameters, the total momentum density $\bm\pi$ of the system can be also expressed in 
terms of the two velocities $\bm v_n$ and $\bm v_p$ as
\begin{equation}
\bm \pi  =  
 \,n_n  \bm{p}_n +  n_p \, \bm{p}_{p} = m_n n_n  \bm{v}_n + m_p n_p \bm{v}_p \, ,
\label{mom-density-tot}
\end{equation}
which does not depend explicitly on entrainment. In particular,  once $m_n=m_p$ has been assumed, the azimuthal component is
\begin{equation}
\pi_\varphi  =  
n_n p_{ n \varphi } +  n_p \,  p_{ p \varphi } = m_n x (n_B \Omega_p + n_n \Omega_{np} ) \,.
\label{mom-density-phi}
\end{equation}
The total angular momentum $L$ of the star (directed along the z-axis) is the volume integral of the angular momentum density $x\,\pi_{\varphi}$: 
\begin{equation}
L\, =\,\int x\,\pi_{\varphi} \, d^3\!x
=\,I\,\Omega_p + \Delta L[\Omega_{np}] \, ,  \label{Ltot}
\end{equation}
where
\begin{equation}
I=\frac{8\pi}{3} \int \! dr \, r^4 \, \rho
\label{Itot}
\end{equation}
is the Newtonian moment of inertia of the star
and
\begin{equation}
\Delta L[\Omega_{np}] =\int \! d^3\!x \,(r\sin\theta)^2 \rho_n \, \Omega_{np}
\label{DLnp}
\end{equation}
is the angular momentum reservoir associated with the lag; here,
we also defined the  rest-mass density profiles of the star and of the superfluid component as $\rho=m_n n_B$ and $\rho_n=m_n n_n$,
which are functions of the radial coordinate $r$ only.

As shown in Paper-I, the maximum glitch amplitude corresponding to a given lag $\Omega_{np}$ is
\begin{equation}
\Delta\Omega_p[\Omega_{np}] \, =  \, I^{-1} \, \Delta L [\Omega_{np}] \, = \, \frac{I_n}{I} \, \langle \, \Omega_{np} \, \rangle  \, .
\label{DOnp}
\end{equation}
 To make contact with the terminology used in Paper-I, we introduced the average lag
\begin{equation}
\langle \, \Omega_{np} \, \rangle = I_n^{-1}\Delta L[\Omega_{np}] \, 
\end{equation}
using the Newtonian moment of inertia of the neutron component
\begin{equation}
 I_n=\frac{8\pi}{3} \int \! dr \, r^4 \, \rho_n  \, .
\label{eq:In-new}
\end{equation}
It is thus possible to define the moment of inertia $I_n$ relative to the superfluid reservoir as the normalization factor of the distribution $\Delta L$ defined in 
Eq. \eqref{DLnp}.

Remarkably, the glitch amplitude in Eq. \eqref{DOnp} does not explicitly depend on entrainment: this is not surprising, as it is a direct consequence of 
Eq. \eqref{mom-density-tot}.
The lag  $\Omega_{np}$, however, is a dynamical variable of the model and its time evolution will be affected by entrainment \citep{antonelliCSQCD}. Even if here we will not study the 
equations of motion of the system, we still want to maximize $\Delta\Omega_p$, in order to obtain a theoretical upper limit to the observed glitch amplitudes. 
As explained in Paper-I, this is done by considering the critical lag for unpinning $\Omega^{\rm cr}_{np}$, corresponding to the maximum reservoir that can be sustained by the pinning force
(even if perfect pinning is probably never realized in real neutron stars). 
The upper limit $\Delta \Omega_{\rm max} $ on the glitch amplitude is thus obtained by artificially emptying the whole reservoir of pinned superfluid, namely 
\begin{equation}
\Delta \Omega_{\rm max} \, = \,\Delta \Omega_p [\Omega^{\rm cr}_{np}]  \, .
\label{gl-max-general-np}
\end{equation}

Estimates of $\Omega^{\rm cr}_{np}$ are based on the still poorly known physics of vortices both in the crust and  in the core of a neutron star.
For this reason, here we construct the critical lag for two alternative physical scenarios: when vortex lines have an overall rigidity so that they collectively organize into 
a stable array of paraxial vortex lines, as early suggested by \citet{ruderman74}, and when vortices are slack at the hydrodynamic scale, 
so that any macroscopic portion of superfluid can flow independently from the others. 
In both cases the critical angular velocity lag diverges as $\sim1/x$ near the rotational axis. 
This, however, is not a flaw of the model: firstly, we do not expect the reservoir to be completely filled
in real neutron stars, secondly, this kind of divergence is cured by the fact that near the rotational axis the moment of inertia density 
on a meridional slice of the star (the $x$-$z$ plane) goes to zero as $\sim x^3$.
\newline
\newline
\emph{Rigid vortex lines - }  This scenario is the one discussed in Paper-I, where the infinite rigidity assumption is nothing but a 
prescription used to model the configuration dynamics of vortex lines: 
an axially symmetric and rigid configuration of vortices is a one-parameter family of curves $\gamma_x$ that foliate a meridional slice of the star
(no toroidal vorticity is considered).
Each curve, say e.g. $\gamma_{x_0}$, represents the shape of a vortex line that intersects the equatorial plane at $x=x_0$.
This construction simplifies the dynamical problem in a significant manner, since it is now possible to use just the density of vortex lines on the equatorial plane, a function dependent only on $x$. 
Vortex  creep, if present,  can still modify the density of vortices with the creeping vortex lines continuously rearranging their 
shape from $\gamma_x$ to $\gamma_{x+dx}$ when moving from $x$ to $x+dx$. 

The simplest case of \emph{columnar rotation}, where each curve $\gamma_x$ is parallel to the rotational axis, has been studied in detail in Paper-I;
in this case it is useful to choose as parameter along the curve the coordinate $z$ (the generalization to rigid vortices that bend over macroscopic distances is not difficult to derive). The motivation behind such a scenario is the possibility that an array of quantized lines behaves as a rigid bundle, with tension proportional to the square of the number of vortices, thus providing a realization of the  Taylor-Proudman theorem, as proposed by \cite{ruderman74}.

It is useful to define an auxiliary angular velocity $x \, \Omega_v = p_{n\varphi} /m_n$, namely
\begin{equation}
\Omega_v = \Omega_p + (1-\epsilon_n) \Omega_{np} \,.
\label{omegav-newton}
\end{equation}
Introduction of this quantity is not strictly necessary but can be useful as it is directly related (via the Feynman-Onsager relation, reviewed in Appendix B) 
to the configuration of vortex lines: in particular  $\Omega_v$ is stationary in the ideal limit of perfectly pinned vortices.
This is not true for $\Omega_n$, because entrainment couples the neutron superfluid to the charged constituents of the $p$-component
which undergoes a steady and slow electromagnetic spin-down. 
Depending on the context, we will also use the word ``lag'' to indicate $\Omega_{vp}=\Omega_v - \Omega_p$; we then have the relation
\begin{equation}
\Omega_{vp} =  (1-\epsilon_n) \,  \Omega_{np} \,.
\label{omegav-newton-bis}
\end{equation}
In the idealized scenario of parallel vortex lines, the Feynman-Onsager relation implies that the quantity $\Omega_{vp}$ is columnar, i.e. depends only on the cylindrical radius $x$; conversely, Eq. \eqref{omegav-newton-bis} shows that entrainment makes $\Omega_{np}$ (and $\Omega_{n}$) non-columnar even  in the presence of straight vortices.

The angular momentum reservoir of Eq. \eqref{DLnp} expressed in terms of $\Omega_{vp}$  is 
\begin{equation}
\Delta L[\Omega_{vp}/(1-\epsilon_n)] \, = \,  2\pi \int \! dx \,x^3 \!\int_{\gamma_x} dz \, \frac{\rho_n}{1-\epsilon_n}  \, \Omega_{vp}\, ,
\label{DLvp}
\end{equation}
implying that Eq. \eqref{gl-max-general-np} can be written as
\begin{equation}
\Delta \Omega_{\rm max} \, = \, \frac{I_v}{I} \, \langle \Omega^{\rm cr}_{vp} \rangle  \, .
\label{DOvp}
\end{equation}
Here the moment of inertia $I_v$ is the normalization factor of the distribution $\Delta L$ in Eq. \eqref{DLvp} that acts on the function $\Omega_{vp}$, namely 
\begin{equation}
I_v = \frac{8\pi}{3} \int dr\, r^4 \frac{\rho_n(r)}{1 - \epsilon_n(r)} \,.
\label{eq:Iv_new}
\end{equation}
The unpinning lag $\Omega^{\rm cr}_{vp}$  is found by assuming that the magnitude of the local  Magnus force (per unit length) \citep{epstein_baym92}
\begin{equation}
  |\bm f_M| = \kappa \, \rho_n  \, x \, \Omega_{np} \, = \, \kappa \, \rho_n  \, x \, \Omega_{vp} \, (1-\epsilon_n)^{-1}
  \label{eq:magnus-newtonian}
\end{equation}
integrated along $\gamma_x(z)$ must equal the mesoscopic pinning force (per unit length) $f_P$  integrated along the same curve, namely the \emph{non-local} unpinning condition
\begin{equation}
    \int_{\gamma_x} |\bm{f}_M|=\int_{\gamma_x} f_P 
   \, , 
    \label{eq:global-unpinning}
\end{equation}
as proposed by \citet{pizzochero2011}. In particular, for straight rigid vortices the critical lag for unpinning is also columnar and given by
\begin{equation}
    \Omega^{\rm cr}_{vp}(x) \,=\, 
    \frac{ \int_{\gamma_x}  f_P }{
        \kappa \,x \int_{\gamma_x}  \frac{\rho_n}{1-\epsilon_n} 
        } \,  .
    \label{eq:rigid_condition}
\end{equation}
We remark that the theoretical pinning force $f_P$ and entrainment coefficient $\epsilon_n$ are calculated as a function of the baryon number density (or, equivalently, of the rest-mass density); once the  density profile of the star $\rho=\rho(r)$ has been fixed (by imposing hydrostatic equilibrium for a given mass), one can find the expressions for $f_P(r) $ and $\epsilon_n(r)$.

Since the integral factor in the denominator of $\Omega^{\rm cr}_{vp}$ is the same that appears in Eq. \eqref{DLvp},
it is straightforward to show that Eq. \eqref{DOvp} gives
\begin{equation}
\Delta \Omega_{\rm max}   \, =  \, \frac{\pi^2}{\kappa \,I} \int_0^{R_d} dr \, r^3  \, f_P(r) \, ,
\label{max-gltich-solo-fp}
\end{equation}
where $R_d$ is the radius corresponding to the  interface between the inner and the outer crust (drip radius). This expression was derived  in Paper-I; it shows that, when the maximum reservoir of angular momentum is determined by the unpinning condition on vortices (the ``pinning paradigm''), the maximum glitch amplitude is independent from entrainment: its value depends only on  the radial profile $f_P(r)$ of the pinning force (i.e. on the $f_P(n_B)$ and $\rho(n_B)$ profiles and on the stellar mass). 

As long as only crustal pinning is considered (in this study we assume $f_P(r)=0$ the core, i.e. there is no pinning of vortex lines to magnetic flux tubes, a possibility that we do not consider here), $\Delta \Omega_{\rm max} $ does not depend on whether the vortex lines stretch across the entire neutron star interior (both S- and P-wave superfluidity
reservoir, as discussed in Paper-I) or are limited to the crustal zone (only S-wave superfluidity reservoir, the standard option found in the literature). 
This fact can be understood physically from the relations $\Delta\Omega_p  =  \Delta L / I$ and $\Delta L  = I_n \langle \, \Omega_{np} \, \rangle$; in the scenario where the reservoir is limited to the crust, the S- and P-wave neutron superfluids are weakly interacting and the core superfluid is strongly coupled to the $p$-component; 
thus the moment of inertia $I_n$ refers only to the crustal neutrons (the core ones contributing to the $p$-component) and is strongly reduced with respect to the case of continuous and rigid vortices trough the star. This, however, is compensated by the strongly increased value of the critical lag: while the integrated  pinning force is unchanged, the Magnus force is now integrated along the crust alone and a larger lag is necessary to reach the critical unpinning condition.
\newline
\newline
\emph{Slack vortex lines - } Within our main hypothesis of null macroscopic meridional circulation,
 the maximum glitch $\Delta \Omega_{\rm max}$ is exactly independent on entrainment in two different and opposite physical situations:
 for vortex lines in a paraxial array and for completely slack vortices that only feel the local pinning and the mesoscopic Magnus force.
 In this latter case we assume  the \emph{local} unpinning condition 
\begin{equation}
|\bm{f}_M|=f_P(r) \, ,
\label{eq:local-unpinning}
\end{equation}
so that the critical lag can now be expressed as
\begin{equation*}
\Omega^{\rm cr}_{vp}(x,z) \,=\,\frac{f_P(r) \,(1-\epsilon_n(r))}{\kappa \, \sin\theta \, r \, \rho_n(r) } \, .
\end{equation*}
Substituting in Eq. \eqref{DOvp}, we end up with the same result given in Eq. \eqref{max-gltich-solo-fp}. 

We stress that here ``slack'' is only referred to the behaviour of vortices at the macroscopic hydrodynamic scale: a completely slack vortex (that can stretch at the mesoscopic scale without any energy cost) would bend even over lengths comparable to the radius of the  Wigner-Seitz cells in the crust.
Thus, the slack vortices introduced here are not tensionless at the mesoscopic scale, consistently with the analysis of vortex pinning carried out by \cite{seveso_etal16}, which incorporates the presence of non-zero single-vortex tension to estimate the mesoscopic pinning force per unit length of vortex line. Moreover, differently from the previous case, this scenario of vortices that are macroscopically slack can lead to the development of superfluid turbulence: if vortices pass through the crust-core interface, the non-pinned section of vortex immersed in the core can wrap around the rotational axis in an unstable configuration \citep{greenstein70} and the vorticity can develop toroidal components. It is not currently known, however, whether or not vortex lines continuously pass from the region of $^1S_0$ pairing to the region of pairing in the $^3P_2- {}^3F_2$ channel: studies based on the different superfluid phases of $^3$He seem to indicate a non-trivial behaviour of vortex lines at the phase boundary \citep{finne_3he}.

Note that in the case of slack vortex lines there is no real advantage in preferring $\Omega_v$ instead of $\Omega_n$, since there is no need to consider a particular geometry of the vortex configuration: the same line of reasoning can be followed by using  directly Eq. \eqref{DOnp} and $\Omega^{\rm cr}_{np}=\Omega^{\rm cr}_{vp}/(1-\epsilon_n)$.
Moreover, any local phenomenological unpinning condition can be used in place of Eq. \eqref{eq:local-unpinning}. 
For example it could be interesting to replace the usual Magnus force with the Gorter-Mellink form for the mutual friction \citep{Qbook}. 
Since isotropic turbulence is unlikely to be relevant for neutron stars \citep{andersson_turbulence}, here  we take the usual choice of Magnus-like mutual friction; when a better understanding of polarized turbulence in neutron stars is achieved, our model can be easily adapted by choosing a suitable unpinning condition to replace Eq. \eqref{eq:local-unpinning}.
\section{Maximum glitch amplitudes: slow-rotation framework}
\label{sec:GR}
In this Section we generalize the result given in Eq. \eqref{max-gltich-solo-fp} by constructing a model of angular momentum reservoir in general relativity, consistent with pinning of vortex lines to the crustal lattice.
The model is static, meaning that we do not discuss the dynamical equations of the problem (i.e. how 
angular momentum is transferred between the two fluids). Moreover, the stellar structure and composition, as well as the spacetime metric, are treated as a fixed background, calculated by using the slow rotation approximation, where the star is assumed to be rigidly rotating (i.e. the two components are in a state of corotation). 
These are not too severe limitations, since our aim is just to provide an upper limit on the glitch amplitudes and not to describe the rotational dynamics.
However, much of the formalism and some intermediate results reported in this Section can be used as a basis for constructing dynamical 
models of superfluid glitching pulsars (e.g. a slow-rotation general relativistic version of the equations for pulsar rotation proposed in Paper-I). 
An example of dynamical model for pulsar glitches with two rigid components but in full general relativity is discussed in \cite{SC17}, where the
authors study the effect of GR on the characteristic rise time of large glitches.

\subsection{Axisymmetric spacetime}

In depth discussion of axisymmetric spacetime around neutron stars can be found in \citet{FSbook} and references therein. 
Here we just need a few basic notions, following from some strong initial assumptions (that are however standard in the context of isolated and rotating relativistic stars):
the neutron star spacetime is asymptotically flat, stationary and axisymmetric; in particular it is a circular spacetime, 
meaning that there are no meridional macroscopic currents in the fluid.
For the global chart we use Schwarzschild-like coordinates $(t,r,\theta,\varphi)$, such that
the Killing vector associated with stationarity is $\partial_t$ and the circular Killing vector is $\partial_\varphi$.
The metric, see e.g. \citet{hartle_sharp}, can be written in terms of four functions 
$\Phi$, $\Lambda$, $\Xi$ and $\omega$ as (setting natural units $c=1$)
%
\begin{multline}
g = - e^{2 \Phi(r,\theta)} {d}t^2 + e^{2 \Lambda(r,\theta)} {d} r^2 + e^{2 \Xi(r, \theta)} 
\cdot \\ \cdot 
\left[r^2 {d} \theta^2
+ r^2 \sin^2\theta \, ({d} \varphi - \omega(r,\theta){d}t)^2\right] \, .
\label{eq:axialmetric}
\end{multline}
The coordinates $\theta$ and $\varphi$ represent, respectively, the polar and azimuthal angles with respect to the rotational axis of the star (defined as 
the set of points where the circular Killing vector vanishes). 

The study of rotating neutron stars is significantly simplified within the approximation of slow rotation, introduced by \citet{hartle67}: 
for a star with mass $M$ and radius $R$ spinning with an angular velocity $\Omega$, the slow-rotation condition can be written as $R^3 \Omega^2/GM \ll 1$,
which implies the slightly less stringent condition $\Omega R\ll c $ \citep{andersson_comer01}.
It is easy to see that, for a typical pulsar with $M \sim 1 M_{\odot}$ and $R\sim10$ km, spinning at $\Omega \sim 70$ rad/s (like the Vela), this approximation works well. 
The slow-rotation framework is less safe for a millisecond pulsar, but so far only two millisecond pulsars have been seen glitching 
(J1824-2452A, \citealt{cognard_backer04}, and J0613-0200, \citealt{mckee_etal16}) and none of them is in the sample of pulsars studied in \citet{pizzochero17},
due to the small amplitude of their glitches.
\\
Following \citet{hartle67}, at  first order in $\Omega$ the metric in Eq. \eqref{eq:axialmetric} reduces to
\begin{multline}
g = \left[ ( \omega(r) r \sin \theta  )^2 - e^{2 \Phi(r)} \right]  {d}t^2 
+ 
e^{2 \Lambda(r)} {d} r^2 +  r^2 {d} \theta^2 +
\\
+ (r \sin \theta)^2 \left[  {d} \varphi^2  - \omega(r)( d\varphi dt + dt d\varphi) \right] .
\label{eq:hartle_metric}
\end{multline}
Of course, the metric functions that appear here and depend only on $r$ are not  the same of Eq. \eqref{eq:axialmetric}.
The centrifugal force and consequent star deformation appear when second-order corrections in $\Omega$ are taken into account and the spherical structure becomes oblate. 
Within the two-fluid formalism, we work at  first order in $\Omega_p$. Also $\Omega_{np}$ and $\omega$ are considered small; in particular,
the quasi-corotation condition $\Omega_{np}\ll\Omega_p$ can be assumed for the present case of pinning-induced lag.

\subsection{The two-fluid model within the slow-rotation approximation}
The basic two-fluid formalism is briefly reviewed in Appendix A, while an introduction to relativistic rotations can be found in \citet{FSbook}.
Given the metric in Eq. \eqref{eq:axialmetric}, the 3-velocities of the fluids measured by the local zero angular momentum observer (ZAMO) are
\begin{equation}
v_s = r \, \sin\theta \, e^{\Xi-\Phi} (\Omega_s - \omega)
\label{eq:3velAMO}
\end{equation}
where $s \in \{ n,p \}$ is a component label. We impose rigid-body rotation of the $p$-component and a quasi-corotating motion of the $n$-component,
namely $\Omega_n(r,\theta)=\Omega_p+\Omega_{np}(r,\theta)$ and $|r \, \Omega_{np}(r,\theta)|\ll r \,\Omega_p \ll c$
for $r<R_d$ and $\theta \in [0,\pi]$.
The corresponding 4-velocities in the global chart are 
\begin{equation}
u_s = W_s \, e^{-\Phi} \left( \partial_t + \Omega_s \partial_\varphi \right)\,  
\label{eq:4vel}
\end{equation}
where we defined the Lorentz factors 
\begin{equation}
W_s = \left( 1-v_s^2 \right)^{-1/2} .
\label{eq:lorentz}
\end{equation}
The local ZAMO rotates with angular velocity $\omega$ with respect to an observer at rest at infinity; therefore, substitution of $\Omega_s$ with the frame-dragging angular velocity $\omega$ gives the 4-velocity of the ZAMO  in the global chart
\begin{equation}
z = e^{-\Phi} \left( \partial_t + \omega \partial_\varphi \right) \, .
\end{equation} 
This vector field plays a special role in the $3+1$ decomposition of spacetime, since $-z$ is the future-pointing unit normal to 
the 3-surfaces defined by $t= \rm cost$ \citep{FSbook, RZbook}. 

The Komar total angular momentum is given by 
\begin{equation}
L= -\int \left(
T_{\alpha \beta} - \frac{1}{2} T^\nu_\nu g_{\alpha \beta}
\right) (\partial_\varphi)^\alpha \, z^\beta\, dV \, ,
\end{equation}
where  $T^{\mu \nu}$ is the energy-momentum tensor of the system and 
\[
dV= 
e^{\Lambda+2\Xi} \, d^3\!x
\]
is the volume form of the three-surface.
By using the energy-momentum defined in Eq. \eqref{eq:energymomentum2} and remembering that
$W_s = -g(u_s,z)$ and $g(\partial_\varphi,z)=0$, one obtains the expression

\[
L 
= \int\left(W_n p_{n\varphi} n_n + W_p p_{p\varphi} n_p   \right)  e^{\Lambda+2\Xi}\,  d^3\!x \, ,
\]
where $p_{s\varphi}$ are the azimuthal component of the momenta defined in Eqs \eqref{pn} and \eqref{pp}.  
It is tempting to draw an analogy with the non-relativistic relations of Eqs \eqref{mom-density-tot}, \eqref{mom-density-phi} and \eqref{Ltot}.
In particular, we would like to split the azimuthal component of the total canonical momentum density
\footnote{
    Note that, differently from the Newtonian case of Eq. \eqref{mom-density-tot}, the total canonical momentum density in GR turns out to be 
    entrainment dependent.
    Explicit dependence on entrainment, however, appears only at  order $O(\Omega_p \Omega_{np}^2)$. 
    }
\[
\pi_\varphi \, = \, W_n p_{n\varphi} n_n + W_p p_{p\varphi} n_p \, 
\]
as done in Eq. \eqref{mom-density-phi}, in order to single out the superfluid reservoir contribution hidden in $L$.
Due to the presence of the Lorentz factors, however, the relation between the momenta and the velocities in the above equation is 
non-linear, and it is not possible to separate $\pi_\varphi$ into two terms that depend only on $\Omega_p$ and  $\Omega_{np}$ respectively.

We now implement the slow-rotation approximation, by keeping only  terms that are at most linear in $\Omega_p$
and using the metric in Eq. \eqref{eq:hartle_metric}. Direct expansion gives for various quantities of interest 
\begin{align}
& W_s  = 1+O(\Omega_p^2)
\\
& \Gamma =W_n W_p(1-v_n v_p) = 1+O(\Omega_{np}^2)
\\
& \Delta =\frac{v_n-v_p}{1-v_n v_p} = x\, e^{-\Phi}\Omega_{np} + O(\Omega_p^2)
\label{DeltaSlow}
\\ 
& u_s^\varphi =  e^{-\Phi}\Omega_s + O(\Omega_p^3)   
\\
& u_{s\varphi} =  e^{-\Phi}x^2 (\Omega_s-\omega) + O(\Omega_p^3)
\\
& \epsilon_s = \frac{2 \, \alpha}{ n_s \, \mu_s} +O(\Omega_{np}^2) \, ,
\label{slow-rot-approx}
\end{align}
where the Lorentz factor $\Gamma$, the lag velocity in the frame of reference of the normal component $\Delta$, and the entrainment parameters $ \epsilon_s$ are defined in Eqs
\eqref{lor-factor-np}, \eqref{Delta} and \eqref{entrain-par} respectively.
 Using the above approximations in Eqs \eqref{pn} and \eqref{pp}, one obtains the expressions
\begin{align}
& p_{ n \varphi } = \mu_n \, x^2 \,e^{-\Phi}(\bar\Omega_p+(1-\epsilon_n)\Omega_{np}) \, ,
\label{pnphi-hartle}
\\ 
& p_{ p \varphi } = \mu_p \, x^2 \,e^{-\Phi}(\bar\Omega_p+\epsilon_p\Omega_{np}) \, ,
\label{ppphi-hartle}
\end{align}
where $\bar\Omega_p=\Omega_p-\omega$.
Leaving aside the fact that here the canonical momentum is an angular momentum instead of a linear momentum, 
the above equations  are completely analogous to Eqs \eqref{pnphi-newton} and \eqref{ppphi-newton}: 
as expected the mass per baryon $m_n$ is replaced by the enthalpy per baryon $\mu_s$, while the only effect of curved spacetime 
is in the presence of the factor $\sqrt{-g^{tt}}=e^{-\Phi}$. The total angular momentum in the slow-rotation approximation is thus given by
%
\begin{equation}
L = \int \! d^3\!x \, e^{\Lambda-\Phi} \, x^2\, [(\mu_n n_n +\mu_p n_p) \bar\Omega_p + \mu_n n_n \Omega_{np})]\, .
\label{Ltot-rel-no-beta}
\end{equation}
Note that this formula was derived by taking only the linear terms of an expansion in $\Omega_p$ and assuming that $\Omega_n$ is of the same order of $\Omega_p$: for its validity, there is no need to invoke the smallness of the lag expressed by the additional  quasi-corotation condition.

Finally, making use of the Feynman-Onsager relation (whose relativistic generalization is given in Appendix B), we can introduce the auxiliary angular 
velocity $\Omega_v=\Omega_p+\Omega_{vp}$ by imposing that 
\[
g_{\alpha\sigma}u_v^\sigma \, = \, p_{n\alpha}/ \mu_n \, ,
\]
where $u_v$ is the 4-velocity associated with a fictitious $v$-component, defined 
in terms of $\Omega_v$ in the same way as done for $u_n$ and $u_p$, namely via Eqs \eqref{eq:3velAMO} and \eqref{eq:4vel}.
In general, the relation between $\Omega_{vp}$, $\Omega_p$ and $\Omega_{np}$ given above is complicated by
the presence of the Lorentz factors $W_v$ in $u_v$ and $\Gamma$ in $p_n$.
However it can be greatly simplified  by assuming slow rotation: in this case, expansion of Eq. \eqref{pn} and of $g(u_v)$ in 
powers of $\bar\Omega_p$ gives
\begin{equation}
\Omega_{vp} \, = \, ( 1-\epsilon_n )\Omega_{np} +O(\Omega_p^2) \, ,
\label{Ov-slow}
\end{equation}
implying that within the slow-rotation approximation the definition of $\Omega_v$ is  the same  as
that given by   Eq. \eqref{omegav-newton-bis} in the Newtonian framework. 
\subsection{Relativistic corrections to the moments of inertia}
\label{subsec:moment_inertia}
Following \citet{hartle67}, the moment of inertia (to  first order in $\Omega$) for a slowly and rigidly rotating star is
\begin{equation}
I = \frac{8\pi}{3 c^2} \int_0^R dr \ r^4 e^{\Lambda(r)-\Phi(r)} \left(\mathcal{E}(r)+P(r)\right) \frac{\bar{\Omega}(r)}{\Omega},
\label{eq:I_rot}
\end{equation}
where $P(r)$ and $\mathcal E (r)$ are the pressure and  energy density profiles inside the star,  $\bar{\Omega}(r) = \Omega - \omega(r)$ encodes  the rotational frame-dragging in GR and the ratio $\bar{\Omega}(r) / \Omega $ does not depend on the angular velocity $\Omega$ and is smaller than one. Relativistically, the mass density is $\rho=\mathcal E / c^2$, so that we will use mass and energy density  as synonyms.

In the slow-rotation approximation, the background neutron star structure can be found by solving the
spherical Tolman-Oppenheimer-Volkoff (TOV) equations, since deformations are second order in $\Omega$. More precisely,  microscopic calculations (performed at zero temperature in flat spacetime) can provide an equation of state (EOS) for neutron star  matter, namely the dependence  of  energy density, pressure and composition (in particular, the neutron fraction $y_n=n_n/n_B $) on the baryon number density $n_B$, from which a barotropic relation $P=P(\rho)$ can be derived.
Given an EOS and for any fixed mass,  the density profile $\rho(r)$ and the metric  functions   $\Phi(r)$ and $\Lambda(r)$ can obtained through the standard TOV equations, while $\bar\Omega(r)$ follows from the integration of an additional equation. We do not review the method here, since it can be found in many previous articles or books
(see e.g. \citet{GlendenningBook} and references therein). Following this prescription, one can thus find the radial profile of all relevant quantities, in particular $\mathcal E(r)=\rho(r) c^2$, $P(r)$, $n_B(r)$, $y_n(r)$; also the profiles $f_p(r)$ and $\epsilon_n(r)$ can be found from any given microscopic expression for $f_p(n_B)$ and $\epsilon_n(n_B)$.

Although some previous studies have already introduced partial moments of inertia within the slow-rotation approximation 
[as done e.g. by \citet{newton_etal15}], we want to further discuss this issue in the present context, since the derivation of $I_n$ and $I_v$ in a relativistic context is more subtle than in the Newtonian case.
In order to clarify all the assumptions  needed to proceed, we have to come back to the total angular momentum of Eq. \eqref{Ltot-rel-no-beta}.
We first remark  that  the thermodynamic Euler relation for a two-fluid system is \citep{langlois_etal98}
\begin{equation}
\mathcal E = \mu_n n_n + \mu_p n_p  - \Psi \, ,
\label{eq:euler2fluid}
\end{equation}
where $\Psi$ is the generalized pressure of the system.
However, the quantity that plays the role of inertia in Eq. \eqref{eq:I_rot} is the enthalpy density, $\mathcal E+P$, of a barotropic fluid:
we thus impose $\Psi=P$, even if a velocity lag is present (see also Appendix A).  
Now, by looking at Eq. \eqref{Ltot-rel-no-beta}, there is the need to specify $n_n \mu_n$ and $n_p \mu_p$.
We thus impose chemical equilibrium ($\mu_p = \mu_n=\mu^*$), so that the above Euler relation reduces to the enthalpy density of a simple barotropic fluid
\[
\mathcal E + P = \mu^* n_B  \, ,
\]
where $\mu^*$ plays the role of a mean effective inertial mass per baryon \citep{FSbook}. 

In the limit of no differential rotation ($\Psi=P$) and of chemical equilibrium,
we have 
\[
n_n \mu_n = y_n n_B \mu^* = y_n (\mathcal E + P)\, ,
\]
where $y_n=n_n/n_B $ is the superfluid fraction.
Under these two additional assumptions, the total angular momentum in Eq. \eqref{Ltot-rel-no-beta} can be written as
\begin{equation}
L\,=\, I \, \Omega_p \,+\, \Delta L [\Omega_{np}] \, ,
\label{L-rel-splitted}
\end{equation}
where $I$ is given in Eq. \eqref{eq:I_rot} and
\begin{equation}
\Delta L [\Omega_{np}]\,=\,\int \! d^3\!x \, e^{\Lambda-\Phi} \,  y_n(\mathcal E + P) \, x^2 \, \Omega_{np}\, .
\label{DLrel}
\end{equation}
We remark that in the decomposition of the total angular momentum in a  global  component corotating at $\Omega_p$ plus the contribution of the neutron reservoir represented by the lag, as done in Eq. \eqref{L-rel-splitted}, only the global part contains the effect of  frame-dragging: the reservoir $\Delta L [\Omega_{np}]$ presents no factor $\bar{\Omega}(r) / \Omega$. This is not so surprising, since the corrections due to frame-dragging and encoded in the use of   $\bar{\Omega}(r) = \Omega - \omega(r)$ cancel out when considering a lag between angular velocities (cf. also Eqs \eqref{pnphi-hartle} and \eqref{ppphi-hartle}).

Again, we can introduce the partial moment of inertia $I_n$ as the normalization factor of the distribution defined by $\Delta L$;
momentarily reintroducing the $c$ factors, it turns out to be
\begin{equation}
I_n = \frac{8\pi}{3 c^2} \int_0^{R_d} dr \ r^4 e^{-\Phi + \Lambda} \, y_n \left(\mathcal E + P \right) \, .
\label{eq:In}
\end{equation}
This allows to define the average lag  $\langle \, \Omega_{np} \, \rangle $ (weighted with $I_n$) and hence write the angular momentum of the reservoir as  
\begin{equation}
\Delta L [\Omega_{np}] = I_n \langle \, \Omega_{np} \, \rangle \, .
\label{DLrel-bis}
\end{equation}
We point out that, although we used the same symbol, the quantity $I_n$ does \emph{not} represent the moment of inertia $I_n^{\rm tot}$ of the entire $n$-component\footnote{For rigid rotation of the $n$-component, this is given by $I_n^{\rm tot}= \frac{8\pi}{3 c^2} \int_0^R dr \ r^4 e^{\Lambda-\Phi} y_n \left(\mathcal{E}+P\right) \bar{\Omega}_n /\Omega_n$, consistently with Eq. \eqref{eq:I_rot}.}, but only that of the reservoir associated with a given lag, in the sense of Eq. \eqref{DLrel-bis}. While in the Newtonian framework the two quantities are the same, in the relativistic context they are distinguished concepts.

Obviously, this argument introduces the more subtle problem of justifying chemical equilibrium. 
\citet{andersson_comer01} showed that chemical equilibrium between the two components in a neutron star 
implies corotation of them, and it is thus only approximatively realized in our context where the fluids must rotate differentially 
in order to produce a glitch (the slowness of electroweak interactions, however, may help to maintain equilibrium). 
Later, \citet{sourie_etal16} have shown the inverse reasoning: starting from the hypothesis of corotation and assuming chemical equilibrium at the center of the star, 
it is possible to infer chemical equilibrium throughout the entire star. 
 The additional condition of quasi-corotation is then necessary to ensure very small departures from chemical equilibrium and from rigid rotation, and thus guarantee the consistency  of Eq. \eqref{eq:In} with the rigid-body Hartle's formalism.

Finally, from Eq.  \eqref{DLrel} we can derive the moment of inertia of the auxiliary $v$-component. This is trivial in the slow rotation limit because of Eq. \eqref{Ov-slow}: the moment of inertia associated with the $v$-component is thus the relativistic analogue of Eq. \eqref{eq:Iv_new}, namely
\begin{equation}
I_v = \frac{8\pi}{3 c^2} \int_0^{R_d} dr \ r^4 e^{-\Phi + \Lambda}  \,
\frac{y_n \left(\mathcal E + P\right)}{1-\epsilon_n} \, .
\label{eq:Iv}
\end{equation}
A similar definition, but with an additional factor $\bar\Omega/\Omega$, is also present in the work of \citet{newton_etal15}:  in order to account for  entrainment in the crust, the authors simply divide the integrand in Eq. \eqref{eq:I_rot} (limited to the neutron component) by the dimensionless effective neutron mass $m^*_n(r)/m_n=1-\epsilon_n(r)$.  This is not inconsistent with our approach, since in their study the authors are referring to the moment of inertia of the entire (rigid) $n$-component $I_n^{\rm tot}$. 
\subsection{Relativistic corrections to the maximum glitch amplitudes}
In the previous subsection, we derived the relativistic generalization of Eq. \eqref{Ltot} within the slow-rotation framework developed by Hartle;
 the formula for the maximum glitch amplitude derived from angular momentum conservation is still given by Eqs \eqref{gl-max-general-np} and \eqref{DOnp}, 
but now one needs to consider the relativistic definitions of $I$ and $\Delta L$, as given respectively in Eqs \eqref{eq:I_rot} and \eqref{DLrel}. For this calculation it is \emph{not} necessary to introduce explicitly the partial moments of inertia.

Before moving to the numerical estimates of the maximum glitch amplitudes, we still need to discuss the critical unpinning lag in a relativistic framework. As in the Newtonian case, we will evaluate the difference in the maximum glitch amplitudes for two opposite (non-local, or rigid, and local, or slack) unpinning conditions. In both cases, we need the Magnus force per unit length of vortex line; its modulus is given by \citep{langlois_etal98}
\begin{equation}
f_M = \kappa \, m_n n_n v_L \, ,
\label{magnus-rel-modulus}
\end{equation}
where $v_L$ is the speed of a segment of vortex line as seen in the local frame comoving with the superfluid flow.
When lines are pinned, they are forced to move with the normal component and $v_L=\Delta$, the relative speed of the protons 
with respect to the neutrons. By using Eq. \eqref{DeltaSlow}, the local Magnus force is thus
\begin{equation}
f_M = \kappa \, m_n \, y_n \, n_B \, e^{-\Phi} \, x \, \Omega_{np} \, , 
\label{magnus-rel-modulus-slow}
\end{equation}
that is the slow-rotation analogue of Eq. \eqref{magnus-rel-modulus}.
For sake of clarity and in coherence  with Eq \eqref{eq:I_rot}, in the following we reinstate the  $c$ factors in the quantities that will be  evaluated numerically in the next Section (the moments of inertia and the upper limits for the glitch amplitudes).
\newline
\newline
\emph{Rigid vortex lines - } We try to adapt the Newtonian phenomenological treatment of macroscopically rigid vortices to the relativistic context. Not surprisingly, extending to GR the non-local unpinning condition is tricky: to proceed, we somewhat arbitrarily generalize the critical lag of Eq. \eqref{eq:rigid_condition} as
\begin{equation}
\Omega^{\rm cr}_{vp}(x) \, = \, \frac{\int_{\gamma_x} dl \, f_P}{\kappa \, x \int_{\gamma_x} dl  \, \frac{m_n n_n}{1-\epsilon_n}e^{-\Phi}} \,.
\label{eq:omcritrig}
\end{equation}
Our aim at the present level is just to provide a test-case for the non-local unpinning condition: since the actual configuration of vortices in a steadily spinning-down neutron star is unknown (provided that such a stable configuration exists), we decide to parametrize the curves $\gamma_x$ in the planes of constant $t$ and $\varphi$ as $\gamma_{x_0}(z)=(x_0,z)$.
The line element along the curve is 
\footnote{
    In cylindrical coordinates the metric in the Hartle slow-rotation approximation has components
    $g_{xx}= \frac{x^2}{r^2} e^{2 \Lambda} +  \frac{z^2}{r^2}$, $g_{zz}= \frac{z^2}{r^2} e^{2 \Lambda} +  \frac{x^2}{r^2}$ and $g_{xz}= \frac{z x}{r^2} (e^{2 \Lambda} - 1)$.
}
\[
dl=\sqrt{ g_{\alpha \beta}\frac{d\gamma_{x}^\alpha}{dz} \frac{d\gamma_{x}^\beta}{dz} } dz= \sqrt{g_{zz}}dz =
\sqrt{ \frac{z^2}{r^2} e^{2 \Lambda} +  \frac{x^2}{r^2} } dz \,.
\]
Numerically, the critical lag in Eq. \eqref{eq:omcritrig} does not differ significantly from those  presented in Paper-I, with a marked peak in the cylindrical region immersed in the inner crust. 
The corresponding maximum glitch amplitude turns out to be
\begin{multline}
\Delta\Omega_{\rm max} \, =  \,
\frac{4 \pi}{I \kappa}  \int_0^{R_d} \! dx \, x^2 \int_0^{z(x)} \!dz \, 
\frac{y_n\,(\mathcal E + P)}{1-\epsilon_n} \,  e^{\Lambda-\Phi} 
\cdot
\\
\cdot
\left( \int_{\gamma_x} dl \, f_P \right)  
\left( \int_{\gamma_x} dl  \, \frac{m_n n_n c^2}{1-\epsilon_n}e^{-\Phi} \right)^{-1}
\label{gl-max-rigid}
\end{multline}
and is entrainment dependent. This is not a drawback of our choice for the curves $\gamma_x$: the dependence on entrainment cannot be canceled out simply because the integrals containing the rest-mass and enthalpy densities do not simplify, as they do in the Newtonian framework.

Equation \eqref{eq:omcritrig} gives a lag that depends only on $x$ (when Schwarzschild coordinates are used), which, according to the results presented in Appendix B, is realized for bent vortices. Therefore, the assumed straight lines $\gamma_x$ do not follow the lines of macroscopic vorticity, which are expected to bend toward the center of the star \citep{rothen81}.
It may be interesting to study the configuration of bent vortices $\gamma_x$ that are consistent with a columnar $\Omega^{\rm cr}_{vp}(x)$, by using the Feynman-Onsager relation in Appendix B, and then integrate the forces along such curves. This should lead to a slightly smaller maximum glitch amplitude, since vorticity is more concentrated in the central region of the star, and less in the external pinning region, where vortices are completely immersed in the crust \citep{SC17}. In other words, more realistic vortex lines arrangements are expected to make the rigid maximum glitch amplitude more similar to the slack one.
At this level we stick to the assumption of straight lines, just to test the robustness of $\Delta\Omega_{\rm max}$ against different unpinning prescriptions and its dependence on entrainment.
\newline
\newline
\emph{Slack vortex lines -} In the case of slack vortex lines, it is possible to follow a more rigorous line of reasoning. By taking Eq. \eqref{magnus-rel-modulus-slow} and imposing the local condition  of Eq. \eqref{eq:local-unpinning}, we find 
\[
\Omega^{\rm cr}_{np}(r,\theta) = \frac{f_P(r) \, e^{\Phi(r)}}{\kappa \, r \sin \theta  \, m_n n_n(r) } \, .
\]
Then, Eqs \eqref{gl-max-general-np} and \eqref{DLrel} immediately give
\begin{equation}
\Delta\Omega_{\rm max} \, =  \,
\frac{\pi^2}{I \kappa}  \int_0^{R_d} \! dr \,r^3 \, e^{\Lambda(r)} \,\frac{\mathcal E(r) + P(r)}{m_n\, n_B(r)\,c^2} \,f_P(r) \, ,
\label{gl-max-slack}
\end{equation}
in complete analogy with the Newtonian maximum glitch amplitude reported in Eq. \eqref{max-gltich-solo-fp} (in the non-relativistic limit, $P \ll \mathcal E$ and $\rho \approx m_n\, n_B\ $ imply that the fraction in the integral is $\approx 1$). With the local prescription, the effects of entrainment on the maximum glitch cancel out in GR as well.

In the slow-rotation approximation, this critical lag is associated with a relative velocity $\Delta^{\rm cr}=xe^{-\Phi}\Omega^{\rm cr}_{np}$ between protons ad neutrons,
describing a laminar flow on concentric spherical shells.
In particular, by considering that the pinning forces estimated by \cite{seveso_etal16} display peaks of the order $10^{15}\,$dyne/cm around 
$m_n n_B \approx 3\times10^{13}\,$g/cm$^3$, we have
\[
\Delta^{\rm cr} \, = \, \frac{f_P}{\kappa \, m_n \, y_n \, n_B} \, \lesssim \, 10^{-5}  \,c \, .
\] 
For comparison,  the slowest pulsar studied by \cite{pizzochero17} is J0631+1036, whose angular velocity of $21.8\,$ rad/s corresponds to an equatorial velocity 
of  order $ \approx 10^{-3} c $. Thus, the approximations of slow rotation and quasi-corotation are consistent with the pinning paradigm.
\section{Numerical results}
\label{sec:numerical}

In Section \ref{sec:GR}, we generalized the model presented in Paper-I and reviewed in Section \ref{sec:newton}, embedding the previous work in a relativistic framework and proposing 
two different prescriptions to calculate the critical lag for unpinning and the corresponding maximum glitch amplitudes.
We now discuss the numerical results for the partial moments of inertia and the maximum amplitudes; then, by following the simple argument proposed in Paper-I, we estimate $M_{\rm max}$\footnote{In \cite{pizzochero17}, the notation $M_{\rm abs}$ was used instead of $M_{\rm max}$.} , the absolute upper bounds on the mass of a pulsar which were
discussed in \cite{pizzochero17} and for which we want to evaluate the relativistic corrections.

The input used in our numerical calculations is summarized in Table \eqref{tab:nedge}; we adopted three  unified barotropic EOSs (SLy4, BSk20 and BSk21), for which the
superfluid fraction $y_n(n_B)$ is provided together with $P(n_B)$ and $\mathcal E (n_B)$: these are calculated consistently for all regions of the neutron star (hence the adjective ``unified"). For the mesoscopic pinning forces, we used the recent results of \citet{seveso_etal16} (the more realistic case $\beta=3,\,L=5000$; note these pinning forces are given in terms of rest-mass density, i.e. they depend on $n_B$). For entrainment, we adopted the entrainment parameters calculated in \citet{chamel_haensel06} for the core 
and in \citet{chamel12} for the crust (also given in terms  of $n_B$).

\begin{table*}
    \caption{
        We list  some properties of the three EOSs used:  $M_{\text{max}}$ is the  maximum non-rotating gravitational mass
        that the EOS can sustain, while $n_{\rm edge }$ is the baryon density at the crust-core interface 
        [see also \citet{fantina2013} for a study of the global properties of non-rotating neutron stars constructed with the same EOSs used here]. 
        For comparison we  also list the baryon density at which the pinning force used in this work becomes zero.}
    \label{tab:nedge}
        \setlength{\tabcolsep}{15pt}
    \begin{tabular}{@{}cccc@{}}
        \hline
        EoS       & $n_{\rm edge}$ [fm$^{-3}$] &  $M_{\text{max}}$ &           Ref.            \\ \hline \hline
        SLy4      &  0.076$\div$0.077          &   2.05 $M_\odot$   & \citep{douchin_haensel01} \\ \hline
        BSk20     &           0.0854           &   2.16 $M_\odot$  &  \citep{goriely_etal10}   \\ \hline
        BSk21     &           0.0809           &   2.28 $M_\odot$  &  \citep{goriely_etal10}   \\ \hline \hline
        $f_{P}$ &           0.0776           &                     & $\beta=3,\,L=5000$ of  \citet{seveso_etal16}   \\ \hline
    \end{tabular}
\end{table*}
\subsection{Relativistic moments of inertia}
It is well known that the relativistic moment of inertia given in Eq. \eqref{eq:I_rot} can have a marked discrepancy with respect to its non-relativistic counterpart. Although only the total moment of inertia appears in the calculation of the maximum glitch, it is interesting to discuss the relativistic corrections also to the partial ones, since they frequently appear in dynamical studies of pulsar glitches.

A word of caution is necessary, however, regarding the Newtonian framework: since the background configuration is actually
fixed by the integration of the TOV equations, it is not clear what should be interpreted as ``inertia'' of the system in this spurious scenario. In Section \ref{sec:newton}  we adopted a strict Newtonian definition, where inertia corresponds to rest-mass, namely we used the rest-mass density $\rho =m_n n_B$; of course this  is at odds with the relativistic definition of inertia $\rho=\mathcal E / c^2$ of Section \ref{sec:GR}: even in flat spacetime, the energy density provided by the EOS contains the contribution of the hadronic nucleon-nucleon interaction and thus it is in principle a different quantity than  $m_n n_B  c^2$ (numerically, however,  they start to differ  by more than 10\%, only for densities $\gtrsim  10^{15} $g/cm$^3$). Indeed, most studies  existing in the literature take the more consistent choice of always adopting the relativistic definition of density and use it in different prescriptions for the moments of inertia (either the Newtonian expressions or some GR approximations, like the one discussed in \citet{ravenhall_pethick94}). In particular, this was the approach adopted in  
Paper-I as well as in \cite{pizzochero17} and we will adhere to it in the following:  in the Newtonian framework, we 
take $ \rho = \mathcal{E} / c^2 $ and $\rho_n = y_n \rho$ in Eqs \eqref{Itot}, \eqref{eq:In-new} and \eqref{eq:Iv_new}.
Note that the Newtonian maximum glitch amplitude of Eq. \eqref{max-gltich-solo-fp} is not affected by this
alternative choice, as long as one works coherently by using the same definition of the mass density $\rho_n$ also in the Magnus force. This ambiguity is inherent to the spurious nature of the Newtonian scenario  and disappears in the GR framework. 

\begin{figure}
    \centering
    \includegraphics[width=.47 \textwidth]{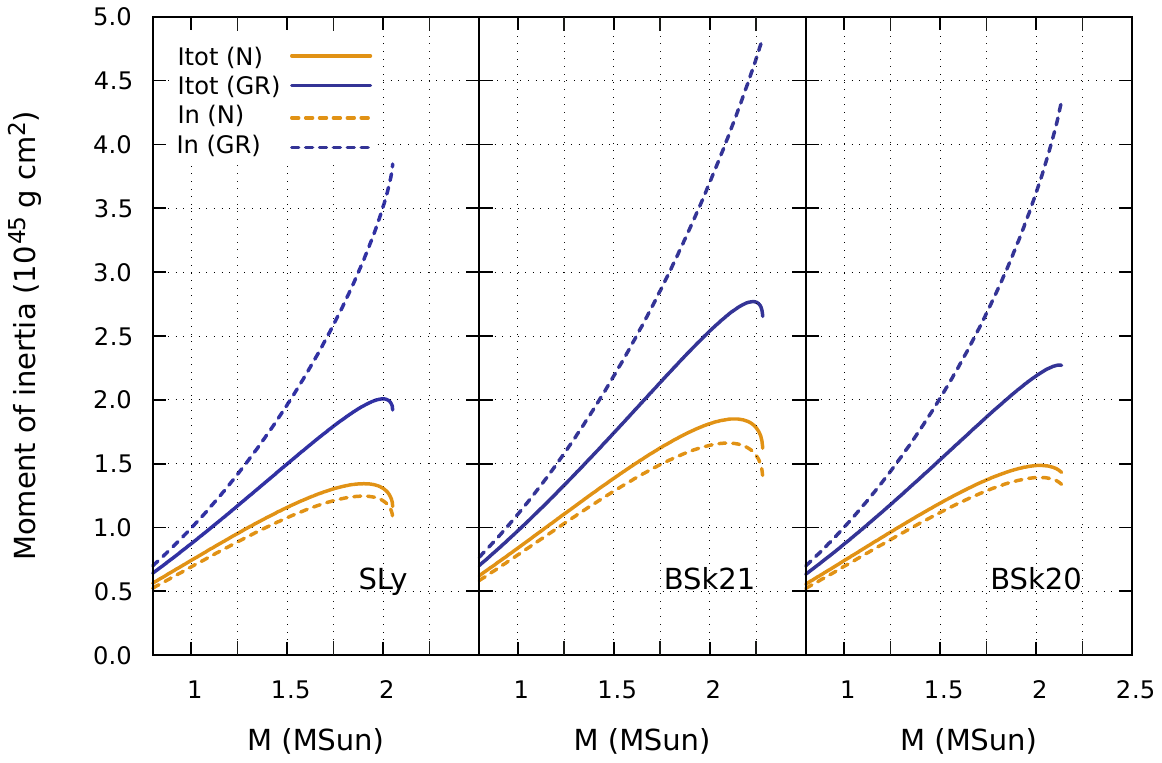}
    \caption{
        The moments of inertia $I$ (solid lines) and $I_n$ (dashed lines) are shown for the three EOSs considered and for the mass interval $[0.8\,M_\odot,2.5\,M_\odot]$. A comparison is made between the non-relativistic moments of inertia (orange curves, labeled by ``N'') and the relativistic ones calculated in the slow-rotation approximation (dark blue curves, labeled by ``GR''). 
        The curves are terminated at the maximum mass allowed by each EOS, as reported in Table \ref{tab:nedge}.
    }
    \label{fig:I_tot}
\end{figure}

In Fig \ref{fig:I_tot} we compare the moments of inertia in the two frameworks (the labels N and GR stand for ``Newtonian'' and ``GR slow-rotation'' respectively) 
by plotting  $I$ and $I_n$ as a function of the gravitational mass $M$ for the three unified EOSs. For the moment of inertia $I_n$ associated with the superfluid reservoir, we have chosen the  scenario of vortices that thread continuously the entire star, so that both the crustal and core superfluid  contribute to the angular momentum reservoir; as already mentioned, this is the scenario investigated in Paper-I and in \cite{pizzochero17}. 

As expected from several existing studies with various EOSs, the total relativistic  moment of inertia is significantly larger than its Newtonian counterpart, with discrepancies up to 50\% for large stellar  masses. The discrepancies are even more dramatic  for the reservoir, where $I_n$ always exceeds the total moment 
of inertia $I$, indicating that the effect of $\bar\Omega/\Omega$ in the integrand of $I$ is more severe than the diminishing effect of $y_n$ in the integrand of $I_n$.  Although unusual this result is not a physical contradiction, as discussed previously. The only consistency requirement is $I_n^{\rm tot} < I $, which holds by construction.

\begin{figure}
    \centering
    \includegraphics[width=.47\textwidth]{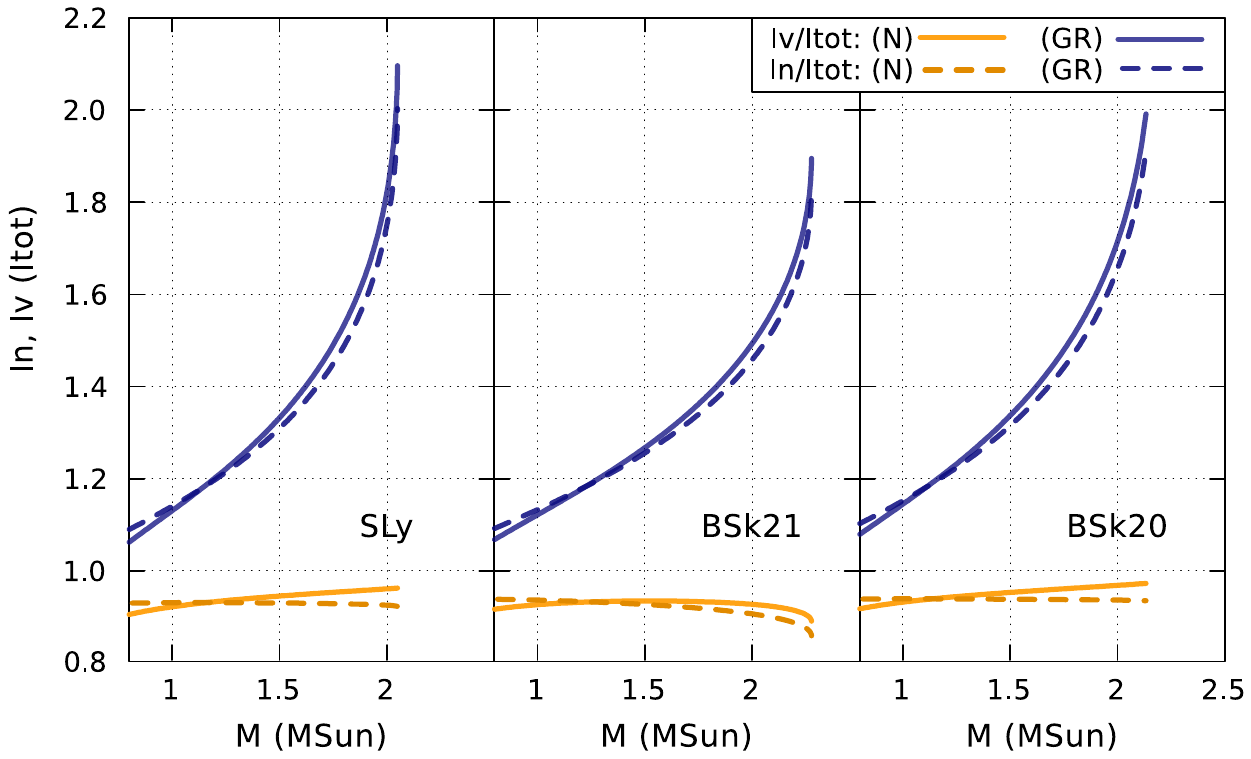}
    \caption{
        Moments of inertia of the superfluid component in the whole star in units of the total moment of inertia  for the mass range $[0.8\,M_\odot,2.5\,M_\odot]$.        
        We make comparison between two cases: when strong entrainment is present (and thus the quantity of interest is $I_v/I$, solid lines) and 
        when the entrainment profile is zero (in this case $I_v$ is equal to $I_n$ and we plot the ratio $I_n/I$, dashed lines). 
        In both cases we show the curves calculated in the Newtonian framework (orange curves, labeled by ``N'') and in the slow-rotation approximation (blue curves, labeled by ``GR'').     }
    \label{fig:In_ratio}
\end{figure}

To better visualize the difference between the two frameworks, in Fig \ref{fig:In_ratio} we plot the ratios $I_n/I$ and $I_v/I$ as a function of mass in the N and GR scenarios for the three EOSs: this kind of ratio is often found in dynamical studies of pulsar glitches. In particular, the figure allows to estimate the influence of entrainment: the advantage of using the $v$-component (determined only by the vortex configuration) is that $I_v$ encodes entirely the effect of entrainment on the physical $n$-component. When entrainment parameters are set to zero $I_v$ tends to $I_n$: therefore, comparison of the two quantities quantifies the global dynamical effect of the non-dissipative coupling between the two components for a given vortex configuration. For the case under study of core plus crust continuous reservoir, the differences are altogether quite small, no more than some percent in the GR scenario. 
Moreover,  for masses larger than $\sim 1.1 \, M_{\odot}$ we find that $I_v > I_n$, while  smaller masses yield $I_v < I_n$, in both frameworks: the entrainment parameters  adopted here are large and negative in the crust, but small and positive in the core. 

\begin{figure}
    \centering
    \includegraphics[width=.47\textwidth]{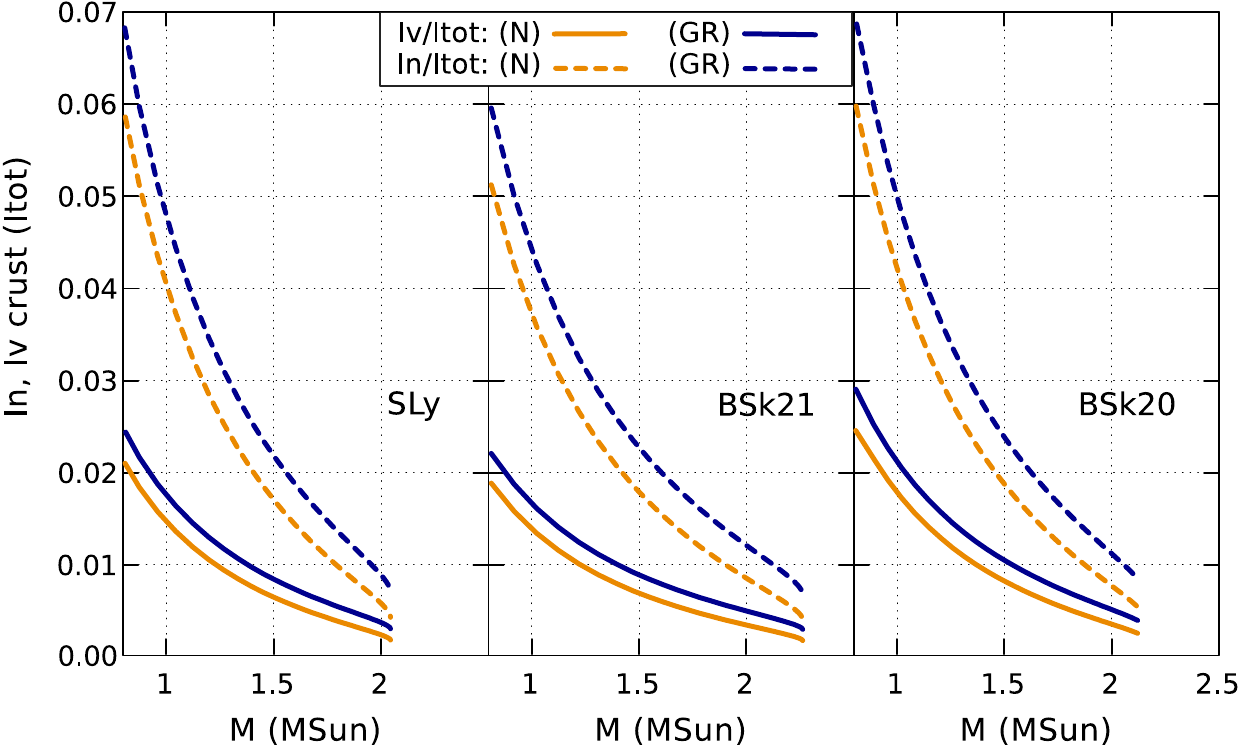}
    \caption{
        Moments of inertia of the superfluid in the crustal pinning region in units of the total moment of inertia for the mass range $[0.8\,M_\odot,2.5\,M_\odot]$; as in Fig \ref{fig:In_ratio}, $I_v/I$ is the case with entrainment (solid lines), while $I_n/I$ is the case without entrainment (dashed lines). Again, we show the curves calculated in the Newtonian framework (orange curves, labeled by ``N'') and in the slow-rotation approximation (blue curves, labeled by ``GR'').
    }
    \label{fig:In_ratio_crust}
\end{figure}

When the superfluid reservoir extends into the core, the crust contribution to $I_v$ dominates for light stars (which present a thick crust), while for more massive stars (with thinner crusts) it is the core contribution that prevails. This is different than the case in which the superfluid reservoir is confined into the crust, defined as the region where $n_B<n_{\mathrm{edge}}$ (see Tab \ref{tab:nedge}): entrainment has a marked decreasing effect on the moment of inertia of the crustal superfluid [cf. \citet{Andersson2012, chamel13}]. As seen in the Fig \ref{fig:In_ratio_crust}, the presence of entrainment actually reduces by a factor 3-4 the effective moment of inertia of the crustal superfluid. On the other hand, the presence of relativistic corrections works in the opposite direction, by slightly increasing $I_v/I$.
\subsection{Maximum glitch amplitudes}
We now come back to the main goal of this paper and we compare the maximum glitch amplitudes  in the two frameworks. Once the input has been fixed (EOS, pinning forces and  entrainment  coefficient),
the maximum glitch amplitude can depend only on the stellar mass. 
In the following, we will discuss three cases, corresponding to the different scenarios explored in the previous Sections:
\newline
\newline
\emph{Model N - }  This is the Newtonian framework  adopted  in \cite{pizzochero17}: the maximum glitch amplitude as a function of mass $\Delta \Omega_{\rm max}(M)$ is calculated with Eq. \eqref{max-gltich-solo-fp}. As already remarked,  the Newtonian result does not depend on the entrainment parameters and it is not necessary to specify how vortices are arranged, since both the parallel  and  slack vortex configurations give the same result. Moreover, also the extension of vortices inside the core is unimportant, as long as vortex lines extend at least 
    across the region where pinning is present. In this paper we assumed the scenario of only crustal pinning and, as reported in Table \ref{tab:nedge}, 
    the region of non-zero pinning lies inside the inner crust for the three EOSs considered. Therefore the Newtonian results for  $\Delta \Omega_{\rm max}(M)$ are valid for both cases of continuous vortex lines and only crustal reservoir.
\newline
\newline
\emph{Model R - } This is the relativistic generalization of  model N for the case of straight rigid vortices, where the non-local unpinning condition is implemented: the function $\Delta \Omega_{\rm max}(M)$ is calculated from Eq. \eqref{gl-max-rigid}.
    In this case the presence of entrainment and the extension of vortices affect $\Delta\Omega_{\rm max}(M)$; the results shown here refer to continuous vortices across the star interior, the general scenario adopted in Paper-I and \cite{pizzochero17}. We remind that  Eq. \eqref{gl-max-rigid} was  actually derived in a non-rigorous way, so that model R should be taken more as a test for the dependence of the maximum glitch amplitudes on phenomenologically reasonable (although not consistent) critical lags, like that of Eq. \eqref{eq:omcritrig}.
\newline
\newline
\emph{Model S - } This is the relativistic generalization of  model N for the case of slack vortices, where the local unpinning condition is implemented: the function $\Delta \Omega_{\rm max}(M)$ is calculated from  Eq. \eqref{gl-max-slack}. This seems to be a natural generalization of its Newtonian counterpart, and all the remarks made for model N are still valid in  this GR extension.
\begin{figure}
    \centering
    \includegraphics[width=.42\textwidth]{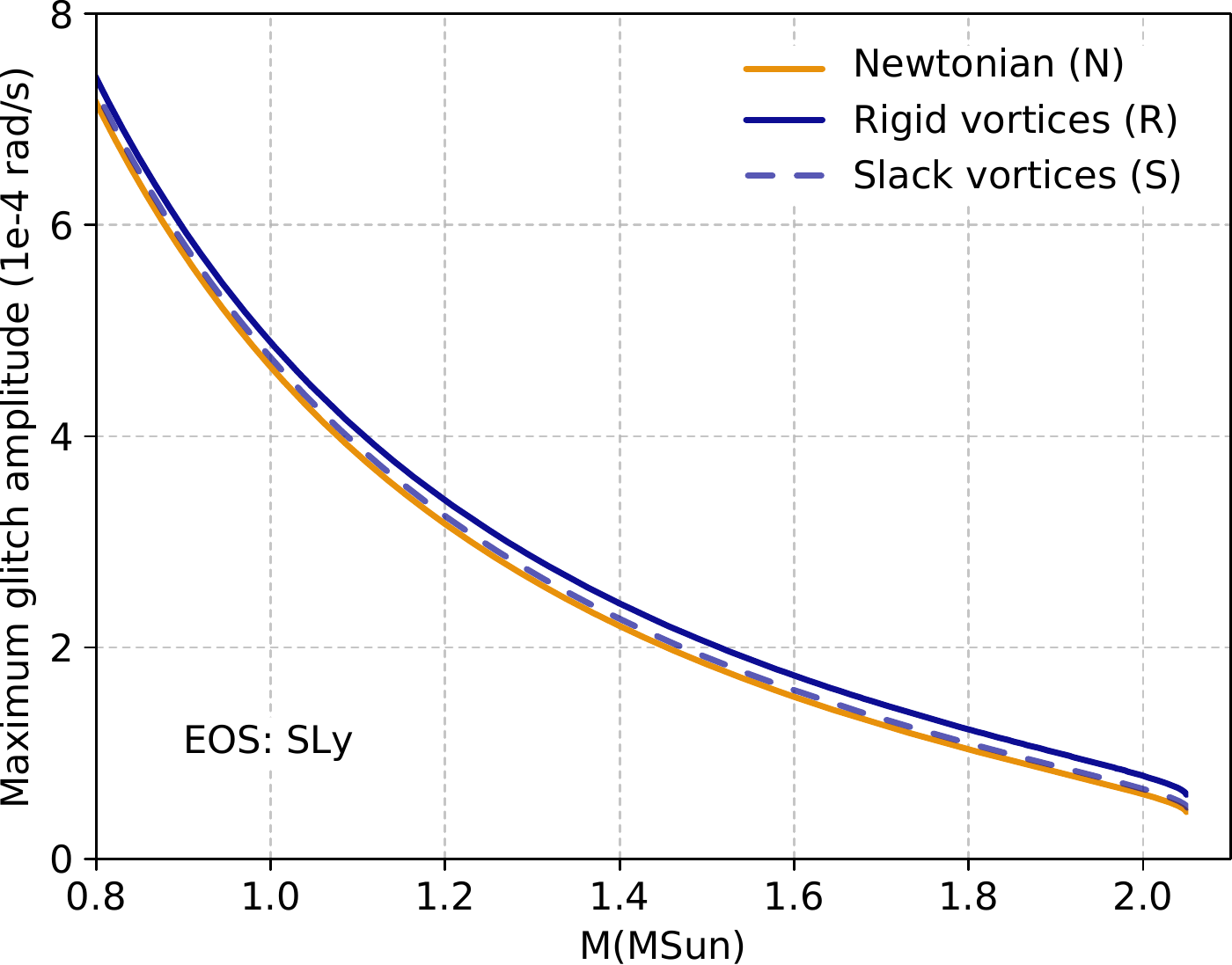}
    \caption{
        The maximum glitch amplitude $\Delta \Omega_{\rm max}$ evaluated with the SLy4 EOS is plotted as a function of mass for the three models studied: the Newtonian case (model N, orange solid line), and the two relativistic cases
        with rigid (model R, blue solid line) and slack (model S, blue dashed line) vortices. For model R, the scenario adopted is that of straight rigid vortices that thread the whole star.
    }
    \label{fig:domega-sly}
\end{figure}
\newline 
\newline
To show an example of the typical result, in Fig \ref{fig:domega-sly} we fix the SLy4 EOS and plot  the function $\Delta\Omega_{\rm max}(M)$ for the three models. We observe that both relativistic models give maximum glitch amplitudes that are slightly larger than their Newtonian counterpart, with model S closer to the non-relativistic case. 

\begin{figure}
    \centering
    \includegraphics[width=.44\textwidth]{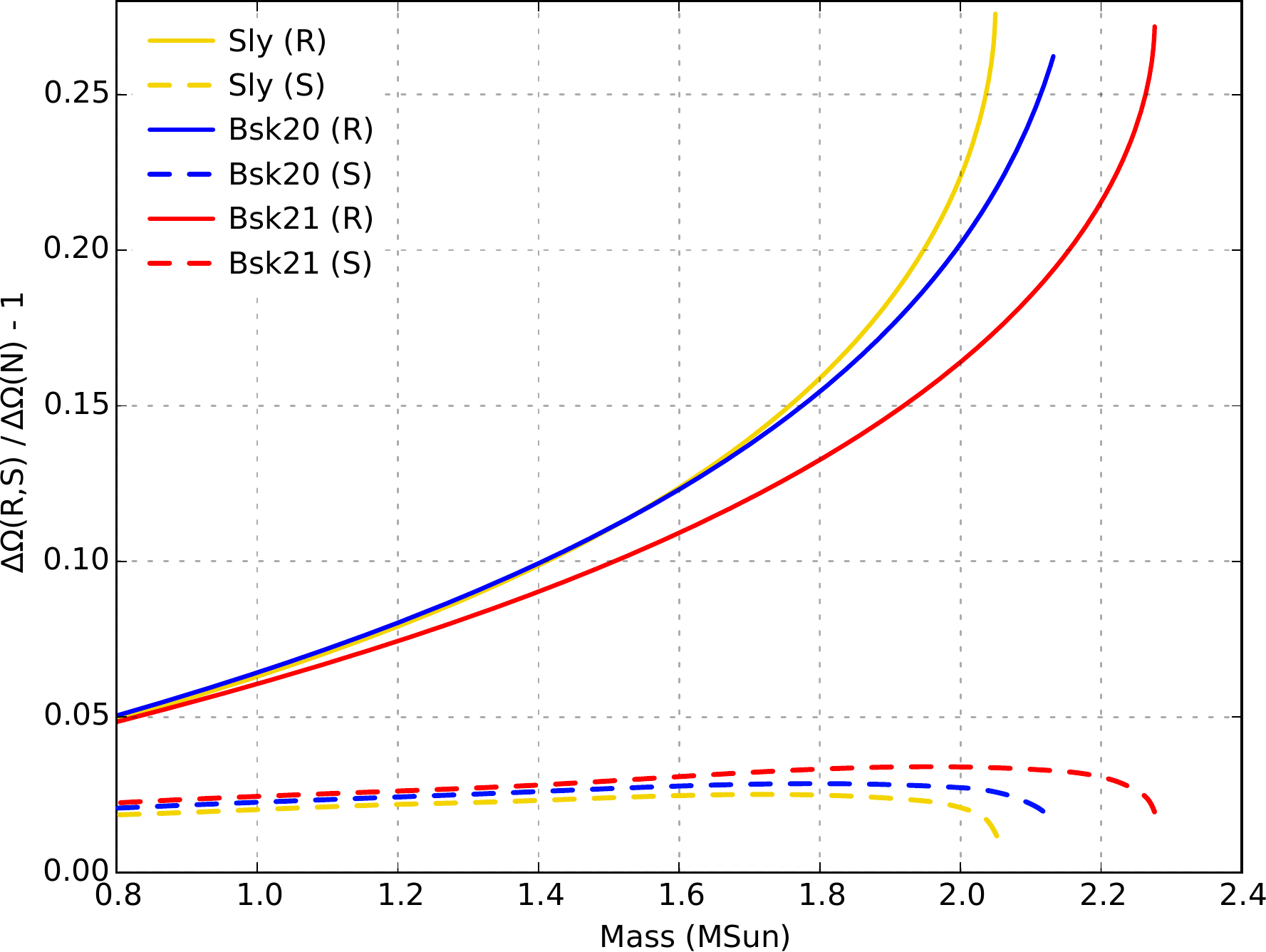}
    \caption{
        Relativistic corrections to the maximum glitch amplitudes $\Delta\Omega_{\rm max}(M)$  for the three EOSs (identified by different colors).  We plot the quantities $\Delta\Omega^{\rm R}_{\rm max}/\Delta\Omega^{\rm N}_{\rm max}-1$ (solid lines) and  $\Delta\Omega^{\rm S}_{\rm max}/\Delta\Omega^{\rm N}_{\rm max}-1$ (dashed lines), where the superscript indicates the model used (cf. Fig \eqref{fig:domega-sly}).
}
    \label{fig:domega-c}
\end{figure}

To better visualize our general results, 
in Fig \ref{fig:domega-c} we show for the three EOSs the relative difference between the relativistic models R and S and the Newtonian one, namely we plot the curves $\Delta\Omega^{\rm R}_{\rm max}/\Delta\Omega^{\rm N}_{\rm max}-1$ and  $\Delta\Omega^{\rm S}_{\rm max}/\Delta\Omega^{\rm N}_{\rm max}-1$, where the superscript indicates the model used. We observe that in model R the relativistic corrections increase with stellar mass, with values  between 5\% and 30\% for all EOSs; conversely, for model S the dependence on mass of the corrections is weak, with values between 3\% and 5\% for all the masses allowed by the EOSs. 

\subsection{Relativistic corrections to the  upper bounds on pulsar masses}
Now, following the argument of Paper-I and using the results of the previous subsection,
we estimate the upper bound on pulsar masses that can be obtained from observations. This method was recently applied to a  sample of large glitchers in  \cite{pizzochero17}:  an upper limit on the stellar mass can be obtained from the largest recorded glitch, while future observations of even larger glitches will further constrain the mass. 
 
\begin{figure}
    \centering
    \includegraphics[width=.47\textwidth]{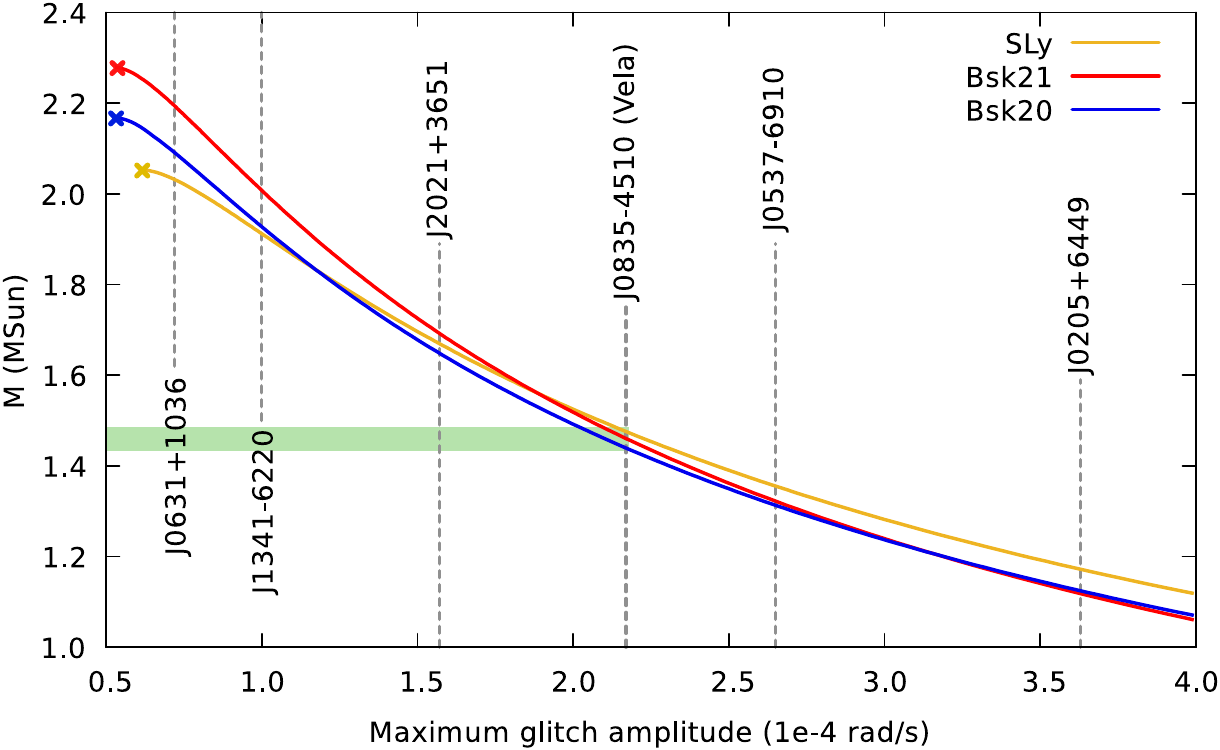}
    \caption{
        Graphical representation of the upper mass limit for a glitching pulsar. 
        In the figure we plot the inverse of the function $\Delta \Omega_{\rm max} (M)$ for the three EOSs; the scenario considered here is that of model R (straight rigid vortices that thread the whole star). We also report the largest observed glitch $\Delta \Omega$ for some of the pulsars
        studied in \citet{pizzochero17}: the errors on  $\Delta \Omega$ are negligible, except for J0537-6910 and J0205+6449, which have a relative error of $\approx 10 \%$.
        For each pulsar, the value of $M_{\rm max}$ is found by considering the intersection of the gray dashed lines (corresponding to the value of $\Delta \Omega$)
      with one of the three curves. Taking the Vela as an example,
        the range of $M_{\rm max}$ arising from the EOSs considered here is highlighted with a shaded band. 
        }
    \label{fig:m-estimate}
\end{figure}

For a given pulsar, whose largest observed glitch amplitude is $\Delta \Omega$, the upper bound on the mass $M_{\rm{max}}$ is given by $\Delta\Omega=\Delta\Omega_{\rm max}(M_{\rm{max}})$. The value of $M_{\rm{max}}$ is only dependent on the choice of the pinning force and the EOS used to calculate the function $\Delta\Omega_{\rm max}(M)$ for models N and S, while model R requires  also the entrainment coefficients (as discussed previously, however, from the results in Fig \eqref{fig:In_ratio}  we  expect the maximum glitch to vary at most  by some percent when entrainment is set to zero).
A graphical representation of the  procedure used to estimate the upper bound is shown in Fig \eqref{fig:m-estimate}, where we plot the inverted function $ M = M(\Delta\Omega_{\rm max}) $ for the three EOSs; here, the curves refer to model R, the one showing the largest relativistic corrections, but  qualitatively these curves are very similar in all models, as can be seen in Fig. \ref{fig:domega-sly}. 
Vertical dotted lines indicate the maximum glitch recorded for a small sample of large glitchers
(the glitch amplitudes are extracted from the Jodrell Bank Glitch Catalog 
\footnote{
    Data are available at \url{www.jb.man.ac.uk/pulsar/glitches.html}, see also \cite{espinoza2011}.
}).

As an example, we consider the benchmark case of the Vela pulsar (J0835-4510), whose largest observed glitch to date has  amplitude $2.17\times10^{-4}\,$ rad/s. By looking at Fig \ref{fig:m-estimate}, the Vela should have a mass lower than $M_{\rm max} \approx 1.5\, M_{\odot}$, when SLy4 or BSk21 are used, slightly less for BSk20. Instead of listing the mass upper bounds corresponding to all the 51 \emph{large} glitchers known to date (those with maximum recorded glitch larger  than $\approx 5\times10^{-5}$ rad/s)  and their deviation with respect to the Newtonian result, we prefer to plot  the discrepancy between the relativistic and non-relativistic  values of $M_{\rm max}$  as a function of the maximum glitch amplitude.

\begin{figure}
    \centering
    \includegraphics[width=.45\textwidth]{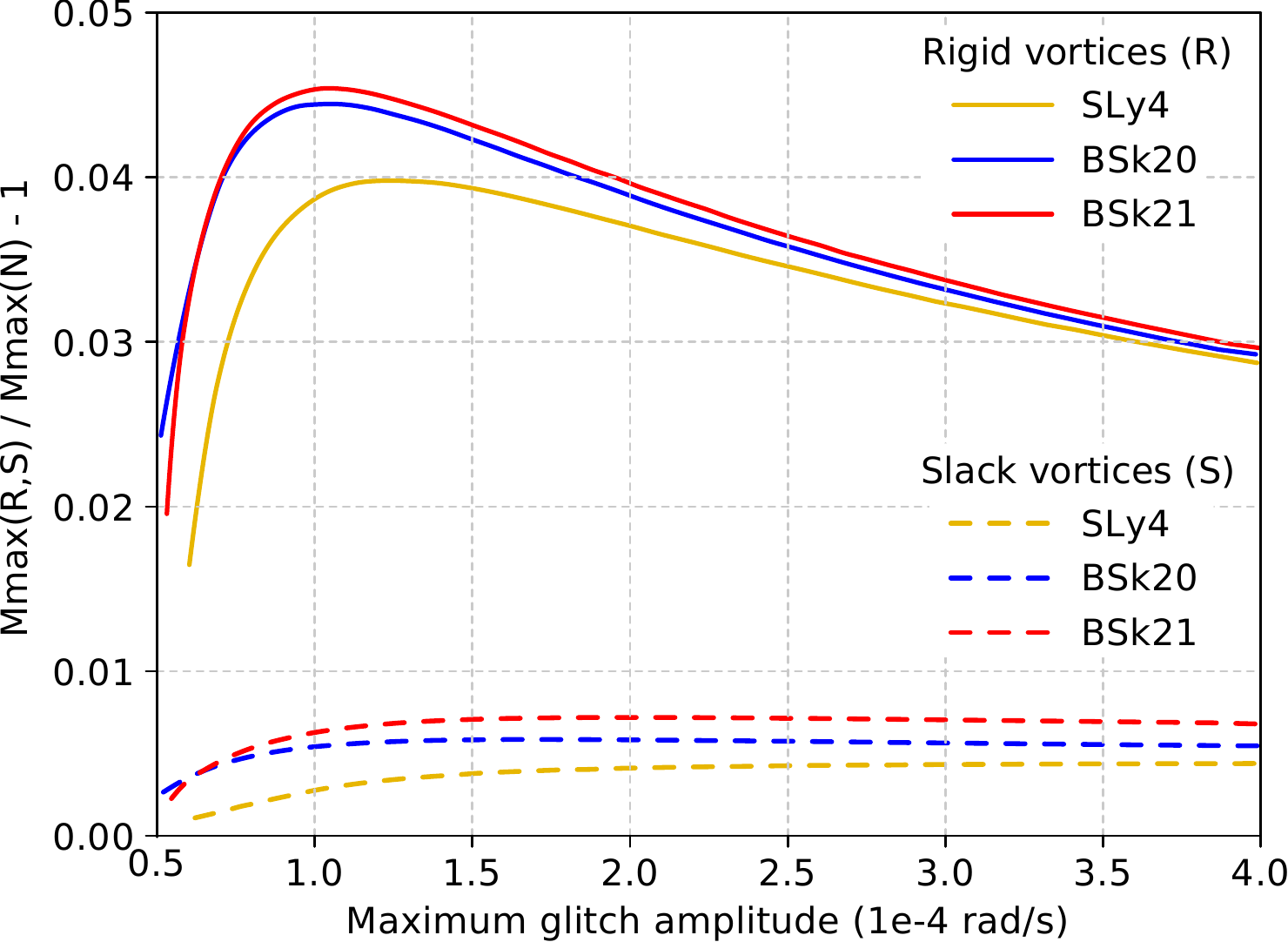}
    \caption{ Relativistic corrections to the mass upper bound $M_{\rm max}$ as a function of the largest observed glitch  for the three EOSs (identified by different colors).  We plot the quantities $M^{\rm R}_{\rm max}/M^{\rm N}_{\rm max}-1$ (solid lines) and  $M^{\rm S}_{\rm max}/M^{\rm N}_{\rm max}-1$ (dashed lines), where the superscript indicates the model used (cf. Fig \eqref{fig:domega-sly}).
                }
    \label{fig:mass-BA-CA}
\end{figure}

In  Fig \ref{fig:mass-BA-CA} we show for the three EOSs the relative difference between the relativistic models R and S and the Newtonian one, namely we plot the curves $M^{\rm R}_{\rm max}/M^{\rm N}_{\rm max}-1$ and  $M^{\rm S}_{\rm max}/M^{\rm N}_{\rm max}-1$, where the superscript indicates the model used. The main remark is that the relativistic corrections to $M_{\rm max}$ are always positive and small, less than 5\% for all masses allowed by the EOSs; in particular, for  model S the discrepancies are smaller than 1\%. The conclusion is that the upper bounds on masses presented in \cite{pizzochero17} are quite robust: in the scenario of slack vortex lines, they are uniquely determined by the pinning force profile and the EOS adopted, while they are independent on entrainment and on the extension of vortices in the core, and are basically unaffected by GR corrections.
\section{Conclusions}
\label{sec:conclusions}

In this article we have generalized the model presented in Paper-I, by embedding it in a relativistic framework and proposing an alternative scenario with slack vortex lines at the hydrodynamical scale. 
We have thus been able to extend to the GR regime the results presented in \citet{pizzochero17} for the upper bounds on pulsar masses that can be derived from the amplitude of their largest observed glitch. 

The line of reasoning is the same for both the Newtonian and the slow-rotation GR formalism: 
the main point is to recognize in the total angular momentum $L$ the contribution of the differentially rotating superfluid, as done in Eqs  \eqref{L-rel-splitted} and \eqref{DLrel}. Then, both the non-local (model R) and local (model S) prescriptions for unpinning allow to calculate the critical lag sustainable by the reservoir of pinned vorticity and from this  the maximum glitch amplitude compatible to the pinning paradigm can be derived. Finally, inversion of the relation $\Delta\Omega_{\rm max}(M)$ enables to determine the mass upper limit as a function of the largest observed glitch.

In the Newtonian framework of Paper-I and \citet{pizzochero17}, here model N, the maximum allowed glitch turns out to be the same for the two vortex scenarios and is given by Eq.  \eqref{max-gltich-solo-fp}: the expression depends only on the pinning force profile and on the EOS chosen to describe neutron star matter. Therefore,  the Newtonian upper bounds on masses are not affected by entrainment or by vortex extension in the core (if  only crustal pinning is considered, as done here).

The slow-rotation GR corrections to the maximum glitch amplitudes depend only slightly on the vortex scenario adopted. The scenario with rigid vortices (model R) gives results for $\Delta\Omega_{\rm max}(M)$ that differ by less than 30\% from their Newtonian counterparts for all the three unified EOSs considered here; moreover, they depend only slightly  on entrainment, as shown by the behaviour of the moments of inertia $I_n$ and $I_v$ of the angular momentum reservoir. Altogether, the upper bounds on masses are increased by less than 5\% when GR effects are accounted for (although not fully consistently in this non-local model).

The scenario with slack vortices (model S) can be extended to GR in a rigorous way, yielding  the expression in Eq. \eqref{gl-max-slack} for the maximum glitch amplitude; this is the natural generalization of its Newtonian counterpart and the same general considerations apply in the present case. We find quite small relativistic corrections  to  $\Delta\Omega_{\rm max}(M)$, of less than 5\%; in turn, this implies that the mass upper bounds are very slightly increased by GR effect, always less than 1\% for all EOSs.

We have also studied the partial moments of inertia in the two extreme scenarios of superfluid reservoir which extends through the whole star and of crust-confined reservoir, presented in Figs \ref{fig:In_ratio} and \ref{fig:In_ratio_crust} respectively. Apart from GR corrections (which tend to increase the ratio $I_v/I$), strong entrainment has much more impact in the case of crust-confined vortices, reducing the ratio $I_v/I$ by a factor of $\sim 3$ (Fig \ref{fig:In_ratio_crust}).

In conclusion, our study shows that the upper bounds on pulsar masses presented in \citet{pizzochero17} are changed by at most some percent when GR is enforced. Altogether, these mass upper limits are robust, since they are not affected by entrainment effects or by superfluid properties of the core-crust interface, both issues being still open at present. Conversely, their dependence on the  pinning force profile and on the EOS describing the whole star can be further explored: different pinning scenarios  (e.g. core pinning) or alternative stellar structures (e.g. exotic interiors) would correspond to different mass constraints. In turn, these could be tested in the future against observations, as more data about glitching pulsars (possibly in binary systems) accumulate.

Our analysis is made simple by the circularity hypothesis, i.e. the macroscopic flow of both components is laminar. However the assumed absence of macroscopic meridional circulation may be in contradiction with the fact that fluid motion in a spinning up (or down) sphere is a combination of a toroidal flow and meridional circulation for all Reynolds numbers, as discussed with applications to neutron stars by \citet{peralta2006} and \citet{vaneysden2013}. In the case of non-zero meridional circulation (i.e. the possible presence of macroscopic toroidal vorticity), the system loses the fundamental invariance under the simultaneous inversion $t \rightarrow -t$, $\varphi \rightarrow -\varphi$, so that in principle the spacetime metric gains additional off-diagonal components.
While this is certainly interesting for a detailed dynamical description of the internal hydrodynamic problem, it is not clear how this further complication would affect the values of global quantities, like the upper bound on the maximum glitch amplitudes discussed in the present work. 

From the point of view of the stellar structure, our model is stationary since we modelled matter as a barotropic fluid at corotation. Within our working hypotheses, the presence of a differential velocity lag of the neutrons with respect to the rigid normal component does not affect the background structural properties of the star (like composition or the frame-dragging angular velocity $\omega$). 
A general formalism to treat slowly rotating superfluid neutron stars can be found in \cite{andersson_comer01}, where the authors determine the effect of the differential rotation of both components on $\omega$, as well as the induced changes in the neutron and proton densities and the change in shape of the star. These refinements are probably worth further investigation and tests with realistic equations of state.


\section*{Appendix A: basic two-fluid formalism}
Our work is based on the standard two-fluid description of neutron stars at zero temperature, wherein all the charged components (protons, electrons and lattice nuclei) are considered as a single normal fluid coexisting with the superfluid neutrons. This is in close analogy with a single-component superfluid at finite temperature, where the normal part represents a fluid of thermal excitations. The magnetic field is assumed to enforce rigid-body rotation.

We adopt the covariant formulation of many interacting fluids developed by Carter and collaborators 
[see \citet{langlois_etal98}, more recent reviews are \citet{AnderssonLivRev} and \citet{chamelReview}], 
where the clear distinction between transport currents and momenta allows a simple implementation of the entrainment effect [see \citet{RZbook} for the distinction between ``coupled'' and ``interacting'' multifluids].  A concise review of the formalism is also presented in \citet{sourie_etal16}.

We consider two fluids, loosely denoted by ``neutrons'' that flow with 4-velocity $u_n^\mu$ and ``protons'' that flow with $u_p^\mu$, both normalized to $-1$. The respective 4-currents and scalar number densities are related to the velocities in the usual way, $n_n^\mu=n_n u_n^\mu$ and $n_p^\mu=n_p u_p^\mu$.
Another fundamental quantity is the Lorentz factor 
\begin{equation}
\Gamma = -u_n^\alpha u_{p\alpha},
\label{lor-factor-np}
\end{equation}
which allows to define the square speed of the neutrons in the frame 
of the protons as
\begin{equation}
\Delta^2 = 1-1/\Gamma^2.
\label{Delta}
\end{equation}   
Note that $\Delta=x|\Omega_n-\Omega_p|$ when we take the non-relativistic limit of our circular model \citep{PC02}.
Once we allow for a non-zero lag $\Delta$ between the components, the system cannot be described by a simple barotropic EOS:
 the internal energy density takes the form $\mathcal{E}(n_p,n_n,\Delta^2)$ with an explicit dependence on $\Delta$.

At this point, it is useful to recall a fundamental property of the single perfect fluid at zero temperature:
the first law of thermodynamics reads $d\mathcal{E} = \mu \, dn_B$, where $n_B$ is the total baryon density and 
$\mu=(P+\mathcal{E})/n_B$ is the chemical potential, that coincides with the enthalpy per baryon. 
The idea is to promote thermodynamically conjugate variables to dynamically conjugate ones by rewriting the first law 
as $-d\mathcal{E} = p_\nu \, dn^\nu$ with the aid of the definition $p_\alpha=\mu u_\alpha$ and of the fact that $dn = -u_\alpha dn^\alpha$ (i.e. $du_\alpha u^\alpha=0$). 
If $-\mathcal{E}$ is interpreted as a Lagrangian density, then $p_\nu$ can be regarded as the canonical momentum per particle.
Similarly, for two interacting fluids we impose that
\begin{equation}
\label{eq:lagrangian}
 -d\mathcal{E} =  p^n_\nu \, dn^\nu_n + p^p_\nu \, dn^\nu_p \, .
\end{equation}

Now the EOS can depend on $\Delta$ and the first law is properly generalized as
\[
d\mathcal{E} = \sum_s \, \mu_s \, dn_s + \alpha \, d\Delta^2 \, ,
\]
where $\mu_s$ are the effective chemical potentials of the two fluids and $s\in\{n,p\}$ is just a label for the two species.
The presence of the scalar function $\alpha$ gives rise to a non-dissipative interaction between the fluids.
Since 
\[
\mu_s dn_s = -\mu_s u_{s \,\alpha} \, dn^\alpha_s
\]
and
\[
d\Delta^2 = 
2\sum_{s\neq q} dn_s^\nu \left( \frac{u_{s\nu}}{\Gamma^2 n_s} - \frac{u_{q\nu}}{\Gamma^3 n_s}  \right) \, ,
\]
it is straightforward to find that 
\[
-d\mathcal{E} = \sum_{s\neq q} dn_s^\nu \left[ u_{s\nu} \left( \mu_s-\frac{2\alpha}{\Gamma^2 n_s} \right) + u_{q\nu} \left( \frac{2\alpha}{\Gamma^3 n_s} \right) \right] \, .
\]
Comparison with \eqref{eq:lagrangian} gives
\begin{align}
p_{n\alpha}/\mu_n & = (1 - \epsilon_n) u_{n\alpha} + (\epsilon_n/\Gamma) u_{p\alpha}  
\label{pn}
\\
p_{p\alpha}/\mu_p & = (1 - \epsilon_p) u_{p\alpha} + (\epsilon_p/\Gamma) u_{n\alpha} \,.  
\label{pp}
\end{align}

Coherently with \citet{RZbook}, \citet{sourie_etal16} and many others, the dimensionless entrainment 
parameters $\epsilon_n$ and $\epsilon_p$ are defined as
\begin{equation}
\epsilon_s = \frac{2 \, \alpha}{\Gamma^2 \, n_s \, \mu_s} 
\label{entrain-par}
\end{equation}
that implies $n_n \mu_n \epsilon_n = n_p \mu_p \epsilon_p$. The same relation
holds in the non-relativistic formulation with the chemical potentials replaced by the masses per baryon $m_s$.
Thanks to this relation between the entrainment parameters, the energy-momentum tensor of the system  
\begin{equation}
T^{\alpha \beta}= n_n p_n^\beta u_n^\alpha + n_p p_p^\beta u_p^\alpha + \Psi\, g^{\alpha \beta}
\label{eq:energymomentum2}
\end{equation}
turns out to be symmetric. 

At corotation, i.e. $\Delta=0$, the two fluids can be in $\beta$-equilibrium ($\mu_n=\mu_p=\mu^*$ and $n_s=n^*_s$). In this case 
$u^\alpha_s =u^\alpha$ and $\Gamma=1$ imply that $p_{s\alpha}=\mu^* \,u_\alpha$, while the energy-momentum tensor 
becomes
\begin{equation}
T^{\alpha \beta}_{(\Delta=0)}= n_B \, \mu^*\, u^\beta u^\alpha + \Psi_{(\Delta=0)}\, g^{\alpha \beta} \, .
\label{eq:energymomentum-1fluid}
\end{equation}
This is the usual energy-momentum form of a single perfect fluid with pressure $P=\Psi_{(\Delta=0)}$
and enthalpy density 
\[
n_B \,\mu^* =\mathcal E ( n_s = n^*_s , \Delta = 0 ) + P \, ,
\]
as can be checked by considering $T^{\alpha \beta}_{(\Delta=0)} u_\alpha u_\beta$.
\section*{ Appendix B: relativistic Feynman-Onsager relation}
The canonical momentum per particle of a perfect fluid at $T=0$ is the 1-form $ p = \mu \, g(u)$ \citep{FSbook},
where $\mu=(\mathcal E +P)/n_B$ is a scalar that represents the enthalpy per particle (or the chemical potential per baryon) and $g(u)$ is the fluid covelocity.
The momentum can thus be naturally integrated over 1-dimensional manifolds embedded in the spacetime.

At the mesoscopic scale the superfluid flow is irrotational, thus the superfluid can rotate only 
if its domain is not simply connected: the topological defects correspond to world sheets into the domain whose 
intersection with the three-space defines a vortex line, see for example \citep{prix2000}.
Integration of $p$ along a closed path $\mathcal C$ inside the superfluid domain is subject to a Bohr-Sommerfeld quantization rule
\begin{equation}
\int_{\mathcal C} p \, = \, \frac{h \, \mathcal N}{2} \, ,
\label{eq:one-fluid-circulation}
\end{equation}
where the factor of $2$ accounts for Cooper pairing, $h$ is the Planck constant and $\mathcal N$
represents the sum of the winding numbers of $\mathcal C$ around each topological defect.

This formula can be made more explicit within the assumption of circular spacetime,
for example by choosing $\mathcal C$ to be an integral curve of $\partial_\varphi$, defined by $t=t_0, r=r_0, \theta=\theta_0$. 
Given the metric in Eq. \eqref{eq:axialmetric}, the azimuthal component of the canonical momentum $p_\varphi$ is actually the angular momentum per baryon 
\[
p_\varphi \, = \,\mu\, W \, e^\Xi r \sin \theta \, v \, = \,\mu\, W\,e^{-\Phi} (e^\Xi r \sin \theta)^2 \bar\Omega \, ,
\]
where $W=(1-v^2)^{-1/2}$ and $v=e^{\Xi-\Phi} r \sin \theta \,\bar\Omega$ is the fluid speed measured by the local ZAMO. 
The angular velocity does not need to be a constant, but can be a function $\Omega(r,\theta)$.
The momentum restricted on the curve $\mathcal C$ is $p|_{\mathcal C}=p_\varphi(r_0,\theta_0) d\varphi$, 
so that the integral in Eq. \eqref{eq:one-fluid-circulation} is trivial and gives 
\[
p_\varphi (r_0,\theta_0)/m_n = \frac{\kappa \mathcal{N}(r_0,\theta_0)}{2\pi} \, .
\]
The bare neutron mass $m_n$ has been introduced in order to obtain the quantum of circulation $\kappa=h/(2m_n)$.
The Feynman-Onsager relation is thus given by 
\begin{equation}
\frac{\bar\Omega}{\sqrt{1-v^2}}\, = \, \frac{\kappa \, e^{\Phi} \, \mathcal{N}(r,\theta)}{2 \pi\, (e^\Xi r \sin\theta)^2(\mu/m_n)} \, .
\label{Fey-rel}
\end{equation}

This relation simplifies within the Hartle slow-rotation approximation: we just have to keep at most the linear terms in $\Omega$
(or $\bar\Omega$)
and use Eq. \eqref{eq:hartle_metric} in order to find
\begin{equation}
\Omega-\omega(r)\, = \, \frac{\kappa \, e^{\Phi(r)} \, \mathcal{N}(r,\theta)}{2 \pi\, (r \sin\theta)^2(\mu(r)/m_n)} \, + \, O(\Omega^2) \, ,
\label{Fey-rel-slow}
\end{equation}
where $\Omega$ is the angular velocity measured by a distant ZAMO. Note that  the Feynman-Onsager relation depends explicitly  on general relativistic frame-dragging, through the metric function $\omega(r)$.

In the Newtonian limit the specific enthalpy is $\mu/(c^2 m_n)\approx 1$
and one recovers the usual non-relativistic Feynman-Onsager relation 
\[
\Omega(r,\theta) \, =\,   \frac{\kappa \, \mathcal{N}(r,\theta)}{2 \pi\, (r \sin\theta)^2} \, .
\]
In this limit, when vortices are parallel to the z-axis, $\mathcal{N}$ is a function of only $r \sin \theta$ and the angular velocity $\Omega$ is columnar.
This ceases to be true even at the level of the slow-rotation approximation due to the presence of the spherical metric functions ($\omega$ and $\Phi$) and of the stratified 
equilibrium enthalpy $\mu(r)$. 
In the above relations, it is also possible to use the fact that the quantity $\bar\mu=\mu(r)e^{\Phi(r)}/m_n$ 
is constant throughout the star \citep{GlendenningBook} 
so that it can be thought as a factor that rescales $\kappa$.

For our two components system,  Eq. \eqref{eq:one-fluid-circulation} is still valid once the four-momentum per baryon $p$ is replaced with $p_{n}$.
This implies that the Feynman-Onsager relation for the two-fluid system with entrainment is obtained by replacing  $\Omega$ with $\Omega_v$ in Eq. \eqref{Fey-rel}.
In particular, for the slow-rotation limit we can  use  Eq. \eqref{Fey-rel-slow} and obtain 
\begin{equation}
\Omega_{vp}(r,\theta)\, = \, -\bar\Omega_p(r) +  
\frac{ \kappa\,e^{2\Phi(r)} \mathcal{N}(r,\theta)}{2 \pi\,\bar\mu_n \,(r \sin\theta)^2}  + O(\Omega_p^2) \, ,
\label{Fey-Ovp}
\end{equation}
where $ \bar\mu_n = \mu_n(r)e^{\Phi(r)}/m_n $ and $\bar\Omega_p(r) = \Omega_p-\omega(r) $.
It is now straightforward to rewrite this equation in terms of $\Omega_{np}$ via Eq. \eqref{Ov-slow} and thus see how entrainment modifies the expression for the physical velocity lag $\Omega_{np}(r,\theta)$  associated with the vortex configuration $\mathcal{N}(r,\theta)$. 

\section*{Acknowledgments}
MA would like to thank Nicolas Chamel and Aur\'{e}lien Sourie for fruitful discussion.
Partial support comes from NewCompStar, COST Action MP1304.
\bibliographystyle{mnras}

\bsp

\label{lastpage}
\end{document}